\PassOptionsToPackage{table,xcdraw,dvipsnames}{xcolor}

\documentclass[twocolumn]{aastex62} 

\usepackage{soul} 
\usepackage{bm} 
\usepackage{amsmath} 
\usepackage{array, makecell, tabularx, cellspace}

\usepackage[table,xcdraw,dvipsnames]{xcolor}
\usepackage{mathrsfs}
\usepackage{hyperref}

\definecolor{blue-violet}{rgb}{0.7, 0.2, 0.8}

\newcommand{\sissa}{SISSA, Via Bonomea 265, 34136 Trieste, Italy \&
INFN Sezione di Trieste}
\newcommand{\ifpu}{IFPU - Institute for Fundamental Physics of the
Universe, Via Beirut 2, 34014 Trieste, Italy}
\newcommand{\iot}{Institute of Optoelectronic Technology, Lishui University, Lishui 323000, China}
\newcommand{\cas}{Xinjiang Astronomical Observatories, Chinese Academy of Sciences, Urumqi 830011, China}
\newcommand{\ioa}{Institute of Astrophysics, Foundation for Research \& Technology -- Hellas (FORTH), GR-70013 Heraklion, Greece}
\newcommand{\mpib}{Max-Planck-Institut f\"{u}r Radioastronomie, Auf dem H\"{u}gel 69, DE-53121 Bonn, Germany}
\newcommand{\icra}{International Centre for Radio Astronomy Research, Curtin University, Bentley, WA 6102, Australia}
\newcommand{\hnu}{Department of Physics and Synergetic Innovation Center for Quantum Effects and Applications, Hunan Normal University, Changsha, Hunan 410081, China}
\newcommand{\hnuu}{Institute of Interdisciplinary Studies, Hunan Normal University, Changsha, Hunan 410081, China}
\newcommand{\LPC}{LPC2E, OSUC, Univ. Orl{\'e}ans, CNRS, CNES, Observatoire de Paris, Universit{\'e} PSL, F-45071 Orl{\'e}ans, France}
\newcommand{\nancay}{Observatoire Radioastronomique de Nan{\c c}ay, Observatoire de Paris, Universit{\'e} PSL, Univ Orl{\'e}eans, CNRS, 
18330 Nan{\c c}ay, France}
\newcommand{\monash}{School of Physics and Astronomy, Monash University, Clayton VIC 3800, Australia}
\newcommand{\ozgrav}{OzGrav: The ARC Center of Excellence for Gravitational Wave Discovery. Clayton VIC 3800, Australia}
\newcommand{\csiro}{CSIRO, Space and Astronomy, PO Box 76, Epping, NSW 1710, Australia}
\newcommand{\astron}{ASTRON, Netherlands Institute for Radio Astronomy, Oude Hoogeveensedijk 4, 7991 PD, Dwingeloo, The Netherlands}
\newcommand{\imapp}{Department of Astrophysics/IMAPP, Radboud University Nijmegen, P.O. Box 9010, 6500 GL Nijmegen, The Netherlands}
\newcommand{\shanghai}{Shanghai Astronomical Observatory, Chinese Academy of Sciences, 80 Nandan Road, Shanghai 200030, China}
\newcommand{\skl}{State Key Laboratory of Radio Astronomy and Technology, A20 Datun Road, Chaoyang District, Beijing, 100101, P. R. China}
\newcommand{\swinburne}{Centre for Astrophysics and Supercomputing, Swinburne University of Technology, Hawthorn, VIC, 3122, Australia}
\newcommand{\inaf}{INAF-Osservatorio Astronomico di Cagliari, via della Scienza 5, 09047, Selargius, Italy}

\graphicspath{{./}{figures/}}

\submitjournal{ApJL}
\received{2025 Sep 02}  
\accepted{2025 Dec 5}   

\shorttitle{Search for GW memory: EPTA, PPTA}
\shortauthors{Tomson et al.}

\begin{document}

\title{Search for Gravitational Wave Memory in PPTA and EPTA Data: A Complete Signal Model} 

\author[0000-0001-7603-1637]{Sharon Mary Tomson}
    \email{sharon.mary.tomson@aei.mpg.de}
\affiliation{Max Planck Institute for Gravitational Physics (Albert Einstein Institute), 30167 Hannover, Germany}
\affiliation{Leibniz Universität Hannover, 30167 Hannover, Germany}

\author[0000-0003-3189-5807]{Boris Goncharov}
    \email{boris.goncharov@me.com}
\affiliation{Max Planck Institute for Gravitational Physics (Albert Einstein Institute), 30167 Hannover, Germany}
\affiliation{Leibniz Universität Hannover, 30167 Hannover, Germany}

\author[0000-0002-6428-2620]{Rutger van Haasteren}
\affiliation{Max Planck Institute for Gravitational Physics (Albert Einstein Institute), 30167 Hannover, Germany}
\affiliation{Leibniz Universität Hannover, 30167 Hannover, Germany}

\author[0000-0002-7176-6690]{Rahul Srinivasan}
\affiliation{\sissa}
\affiliation{\ifpu}

\author[0000-0001-6499-6263]{Enrico Barausse}
\affiliation{\sissa}
\affiliation{\ifpu}

\author[0009-0006-4212-3801]{Yirong Wen}
\affiliation{\cas}

\author[0000-0001-9782-1603]{Jingbo Wang}
\affiliation{\iot}

\author[0000-0003-4453-3776]{John Antoniadis}
\affiliation{\ioa}
\affiliation{\mpib}
\author[0000-0002-8383-5059]{N. D. Ramesh Bhat}
\affiliation{\icra}
\author[0000-0001-7016-9934]{Zu-Cheng Chen}
\affiliation{\hnu}
\affiliation{\hnuu}
\author[0000-0002-1775-9692]{Ismael Cognard}
\affiliation{\LPC}
\affiliation{\nancay}
\author[0000-0003-3432-0494]{Valentina Di Marco}
\affiliation{\monash}
\affiliation{\ozgrav}
\affiliation{\csiro}
\author[0000-0002-3407-8071]{Huanchen Hu}
\affiliation{\mpib}
\author[0000-0003-3068-3677]{Gemma H. Janssen}
\affiliation{\astron}
\affiliation{\imapp}
\author[0000-0002-4175-2271]{Michael Kramer}
\affiliation{\mpib}
\author[0009-0009-9142-6608]{Wenhua Ling}
\affiliation{\csiro}
\author[0000-0002-2953-7376]{Kuo Liu}
\affiliation{\shanghai}
\affiliation{\skl}
\affiliation{\mpib}
\author[0009-0001-5633-3512]{Saurav Mishra}
\affiliation{\swinburne}
\affiliation{\ozgrav}
\affiliation{\csiro}
\author[0000-0002-1806-2483]{Delphine Perrodin}
\affiliation{\inaf}
\author[0000-0001-5902-3731]{Andrea Possenti}
\affiliation{\inaf}
\author[0000-0002-1942-7296]{Christopher J. Russell}
\affiliation{\csiro}
\author[0000-0002-7285-6348]{Ryan M. Shannon}
\affiliation{\swinburne}
\affiliation{\ozgrav}

\author[0000-0002-3649-276X]{Gilles Theureau}
\affiliation{\LPC}
\affiliation{\nancay}

\author[0000-0003-4498-6070]{Shuangqiang Wang}
\affiliation{\cas}
\affiliation{\csiro}

\begin{abstract}

We perform searches for gravitational wave memory in the data of two major Pulsar Timing Array (PTA) experiments located in Europe and Australia. 
Supermassive black hole binaries (SMBHBs) are the primary sources of gravitational waves in PTA experiments. 
We develop and carry out the first search for late inspirals and mergers of these sources based on full numerical relativity waveforms with null (nonlinear) gravitational wave memory. Additionally, we search for generic bursts of null gravitational wave memory, exploring possibilities of reducing the computational cost of these searches through kernel density and normalizing flow approximation of the posteriors. 
We rule out the mergers of SMBHBs with a chirp mass of $10^{10}~M_\odot$ up to $700$~Mpc over $18$ years of observation at $95\%$ credibility. 
We rule out the observation of generic displacement memory bursts with strain amplitudes $> 10^{-14}$ in brief periods of the observation time but across the sky, or over the whole observation time but for certain preferred sky positions, at $95\%$ credibility.

\end{abstract}

\keywords{gravitational waves --- 
pulsars: general --- methods: data analysis}

\section{Introduction} \label{sec:intro}

Pulsar Timing Array~\citep[PTA,][]{FosterBacker1990} experiments have a primary goal of detecting nanohertz-frequency gravitational waves through decade-long observations of pulse arrival times from a set of millisecond pulsars~\citep{Sazhin1978,Detweiler1979}. 
This goal is now within reach, with various lines of evidence for the stochastic background in the data of all major PTAs, albeit with varying and sometimes debatable levels of statistical significance~\citep{NG_12_GWB,PPTA_DR2_GWB,EPTA_CRN,IPTA_DR2_GWB,NG_15_GWB,EPTA_DR2_GWB,PPTA_DR3_GWB,CPTA_DR1_GWB,MT_DR1_GWB}. 
Gravitational waves are sourced from a mass quadrupole moment, resulting in the primary oscillatory strain time series $h(t)$ observed at detectors. 
It is expected that the superposition of many such mass quadrupole sources, including supermassive black hole binaries (SMBHBs), constitutes the background. 
With this, General Relativity (GR) predicts that the emission of gravitational waves is always accompanied by an additional, non-oscillatory contribution to the signal known as the gravitational wave memory~\citep{Favata2010};  the standard position-shift effect is often termed \emph{displacement memory}~\citep{StromingerZhiboedov2014}. 

Gravitational wave displacement memory is the net change in the gravitational wave strain caused by the propagation of the mass-energy from the gravitational wave source to infinity. 
The ordinary (linear) memory was discovered by~\citet{Zel'dovichPolnarev1974} when considering unbound masses (example - unbound black holes in hyperbolic trajectories, gamma ray bursts or supernovae events). 
The null (nonlinear)\footnote{When referring to types of memory, ``ordinary'' and ``null'' are the modern terms, whereas ``linear'' and ``nonlinear'', respectively, are the legacy terms \citep{ MitmanMoxon2020, MitmanBoyle2024, Bieri2021}.} memory was discovered by~\citet{Christodoulou1991} by considering compact binaries.
In this case, the unbound energy is carried out by gravitational\footnote{In the most general case, also electromagnetic radiation~\citep{BieriGarfinkle2013}.} radiation to null infinity, as suggested by~\citet{Thorne1992}. This null memory is a nonlinear effect in GR due to being sourced by the backreaction of gravitational waves onto the stress energy tensor.

In addition to the displacement memory, there are higher-order memory terms associated with higher-order temporal moments of the curve deviation near null infinity~\citep{GrantMitman2024}. 
These include, the spin memory~\citep{PasterskiStrominger2016} and the center-of-mass memory~\citep{Nichols2018}. 
It is worth emphasizing that the gravitational wave memory is tied to symmetries of spacetime through Noether's theorem. 
Other effects previously termed ``memory'' unrelated to symmetries, such as ``velocity memory'', are now referred to as ``persistent observables''~\citep{FlanaganGrant2019}. 
Fascinatingly, memory effects, spacetime symmetries, and soft theorems~\citep{Weinberg1964}, derived in different formalisms spanning GR and quantum field theory over decades of studies, are found to represent the same physical phenomenon.
This connection is known as the infrared triangle~\citep{StromingerZhiboedov2014}. 
Therefore, the detection of gravitational wave memory will enable studies of spacetime symmetries and tests of the nonlinear nature of GR~\citep{GoncharovDonnay2024}. 

Detecting a nonlinear displacement memory signal with PTAs depends on the occurrence of mergers of SMBHBs. While these systems are the dominant sources of the stochastic gravitational wave background in the nanohertz band, individual mergers which can produce detectable memory are rare. Moreover, the rates of such mergers are not well constrained, introducing significant uncertainty in the expected number of detectable memory events.
Observing the scattering of SMBHBs producing an ordinary memory is much less certain, as it is not clear how the scattered black hole would escape to asymptotic infinity in the vicinity of a galactic center. 
Luckily, observing spans of contemporary PTAs have already exceeded a few decades, while detector sensitivity continues to improve~\citep{HobbsManchester2020,ChenXu2025,SpiewakBailes2022}. 
Memory bursts were first studied theoretically in the PTA context by several early works. \citet{Seto2009} outlined how pulsar timing arrays could detect both inspiral and memory components from SMBHBs. Near-simultaneously, \citet{vanHaasterenLevin2010} developed more detailed signal modeling to explore detection prospects of memory events followed by \citet{PshirkovBaskaran2010}, who assessed PTA sensitivity to such events.
In all of these studies, the net strain displacement was modeled as a step function. 
We refer to this approximation as the \textit{memory burst}.
The detectability of this signal with an array of $20$ pulsars observed with a precision of $100~\text{ns}$ and approximately bi-weekly cadence over the period of $10$ years ($250$ pulse arrival time measurements) was reported by~\citet{vanHaasterenLevin2010}.
They found, with simplified noise assumptions, that the black hole merger with $\mathcal{M}=10^{8}~M_\odot$ at $z=0.1$ (equivalently, $\mathcal{M}=10^{9}~M_\odot$ at $z=1$) can be detected with $2\sigma$ confidence. 
\citet{EnokiInoue2004} expect an event rate of SMBHB mergers with a total mass of $M\sim 10^{8}~M_{\odot}$ at redshifts $z<1$ to be $0.1~\rm events~yr^{-1}$. 
Overall, \citet{vanHaasterenLevin2010} estimate $\sim 0.01 - 0.1$ detectable memory bursts over the $10$-year observation time. 
\citet{CordesJenet2012} presents the same estimate, suggesting a possible improvement when considering memory bursts at pulsar positions. More recent population studies find very low PTA rates for detectable memory. 
For example, ~\citet{RaviWyithe2015} predict that over a 10\,yr PTA span only $\sim 10^{-5}$ bursts with memory amplitude $>5\times10^{-15}$ and $\sim 10^{-3}$ bursts with $>2\times10^{-15}$ amplitude would occur.
\citet{IsloSimon2019} forecast that memory bursts with a final gravitational wave strain offset of atleast $5\times10^{-16}$ occur only $8.7\times10^{-4}$– $5.2\times10^{-3}$ times per century with a more optimistic model.

PTAs have performed several searches for displacement memory. 
The Parkes Pulsar Timing Array (PPTA), using the frequentist approach, performed a search for memory bursts in their initial dataset, setting an upper limit on the rates of such events which could be detected \citep{WangHobbs2015}. 
The North American Nanohertz Observatory for Gravitational Waves (NANOGrav) conducted searches on their 11-year \citep{NG_11_MEM}, 12.5-year \citep{NG_12_MEM}, and 15-year datasets~\citep{NG_15_MEM}, using the Bayesian approach to data analysis. 
These analyses have found no statistically significant evidence for memory signals, with the 15-year dataset yielding a Bayes factor of $3.1$ ($\ln \mathcal{B} = 1.13$) in favor of a model including memory burst against the null hypothesis including pulsar-intrinsic noise and common time-correlated stochastic signal. 
Upper limits have been placed on the strain amplitude as a function of sky location and burst time. 
\citet{DandapatSusobhanan2024} searched for ordinary displacement memory from hyperbolic SMBHB encounters using the NANOGrav 12.5-year data. To date, no search for gravitational wave memory bursts has been performed using EPTA data. 
In contrast~\citet{WangHobbs2015} conducted a search on PPTA data, but that was $10$ years ago. 
Thus, it is of interest to perform searches on the latest EPTA and PPTA data. 

There is a motivation to improve searches for gravitational wave memory with PTAs. 
First, the memory burst approximation is suboptimal for SMBHBs, which are the dominant sources of nanohertz gravitational waves. A full waveform model for an SMBHB merger consists of the inspiral, merger, ringdown, and memory components.
Contemporary continuous gravitational wave searches from SMBHBs with PTAs only look for the inspiral components, assuming the signal to be monochromatic on the observational timescales of decades. 
The search for ordinary memory from hyperbolic SMBHBs by \citet{DandapatSusobhanan2024} is beyond a simple burst, but a search for mergers with the most abundant bound SMBHBs producing null memory has not been performed yet. 

Second, searches for gravitational wave memory with PTAs are becoming computationally expensive due to the ever-increasing complexity of data and noise, as well as the necessity to model the gravitational wave background. \citet{SunBaker2023} introduce the Factorized Bayesian approach based on likelihood factorization and the use of lookup tables. 
This method has been used in~\citet{NG_12_MEM,NG_15_MEM}. 
However, lookup tables are a discrete approximation of the likelihood. 
Modern data analysis tools allow for the introduction of smooth and continuous approximations of the likelihood.

In this work, we address the above limitations and opportunities.
We develop the first methodology to search for SMBHB mergers using the full waveform from numerical relativity, including the inspiral phase, the merger, the ringdown, and the memory in contrast to the inspiral only waveforms employed in continuous wave searches \citep{TomsonGoncharov2025}. 
We also develop a methodology to search for gravitational wave memory using likelihood factorization with kernel density estimation~\citep[KDE][]{} and normalizing flows~\citep{SrinivasanCrisostomi2024}. 
Next, we perform searches for SMBHB mergers, and generic gravitational wave memory bursts in the second data release of the European Pulsar Timing Array (EPTA DR2) and the third data release of the Parkes Pulsar Timing Array (PPTA DR3).

The rest of the paper is organized as follows. 
In Section~\ref{sec:data}, we describe properties of the data used in this work, EPTA DR2 and PPTA DR3. 
In Section~\ref{sec:method}, we outline the methodology of our searches. 
In Section~\ref{sec:results}, we present our results. 
In Section~\ref{sec:discussion}, we highlight important aspects of memory searches to keep in mind for future work. 
We summarize the conclusions for gravitational wave memory in Section~\ref{sec:conclusion}.

\section{\label{sec:data}The datasets}

The Parkes Pulsar Timing Array third data release~\citep[PPTA DR3,][]{PPTA_DR3_TIMING} contains observations of 32 pulsars over 18 years with a cadence of 3 weeks. 
The data was obtained using the 64-m Parkes `Murriyang' radio telescope. From its onset in 2004, this telescope has been continuously acquiring data from 48 hours of observations every 2-3 weeks.
PPTA DR3 is a combination of the data from PPTA's previous release (data up to 2018) along with data acquired from an ultra-wideband low-frequency receiver (UWL)~\citep{HobbsManchester2020}. 
The UWL receiver observes in a continuous radio frequency bandwidth between 704 and 4032 MHz, enhancing the instantaneous sensitivity. 
Narrow-band pulsar timing is implemented, and the arrival times are acquired using a frequency-dependent pulse portrait.

The European Pulsar Timing Array second data release~\citep[EPTA DR2,][]{EPTA_DR2_TIMING} contains data from 25 pulsars which were collected from six European radio telescopes: the Effelsberg 100-m radio telescope, the 76-m Lovell Telescope, radio telescope at Nan\c{c}ay Radio Observatory, the 64-m Sardinia Radio Telescope and the Westerbork Synthesis Radio Telescope. 
Compared to the previous data release, which only included data from the legacy data recording system, EPTA DR2 adds data from next-generation coherent de-dispersion recording systems covering a wider bandwidth and thus better sensitivity at each telescope. Time spans for pulsar datasets range from 14 to a maximum of 25 years with a broad frequency coverage between about 300~MHz and 4~GHz. 

We perform our analysis with two versions of EPTA DR2: the complete dataset and a shortened version that includes only the observations made with the next generation of backend-receiver combinations. 
These newer backend systems use field-programmable gate array (FPGA) hardware to digitize the incoming electrical signals. 
The complete dataset, covering 24.7 years of observation, will be referred to as the \textit{EPTA 25-year data}.
The truncated version, including only the last 10.3 years of observation collected using newer wideband backends, referred to as the \textit{EPTA 10-year data} henceforth.

Our pulsar-specific noise models are as follows.
For EPTA DR2, our models are based on the result of the analysis by~\citet{GoncharovSardana2024}, excluding hierarchical modeling of red noise. 
\citet{GoncharovSardana2024} builds their noise models on top of 
the original EPTA DR2 noise analysis by the~\citet{EPTA_DR2_NOISE}. 
For PPTA DR3, we adopt the noise models from~\citet{PPTA_DR3_NOISE}. 

We model the contribution of the gravitational wave background as temporally correlated and spatially uncorrelated (\textit{i.e.}, no Hellings-Downs correlations) stochastic process, with the same power spectrum of temporal correlations across pulsars. This simplification enables significantly faster Bayesian inference by reducing the complexity of the covariance matrix. The procedure is justified because a null memory signal is deterministic, and hence not intrinsically correlated with the GWB spatial covariance structure.  However, in the case of the EPTA 10-year dataset, we include HD correlations in the model, as preliminary analyses showed a spurious signal when spatial correlations were unaccounted for. Further details and discussion are provided in the results section.

\section{\label{sec:method}Methodology}

\subsection{Bayesian inference}
\label{sec:method:bayesian}

We use the standard PTA Gaussian likelihood for the timing–residual data vector $\bm{\delta t}$,

\begin{equation}
    \mathcal{L}(\bm{\delta t} | \bm{\theta}) = \frac{\exp\bigg(  -\frac{1}{2} (\bm{\delta t} - \bm{\mu})^\text{T} \bm{C}^{-1}(\bm{\delta t}-\bm{\mu}) \bigg)}{\sqrt{\det\left( 2\pi \bm{C}\right)}},
\end{equation}

where vector $\bm{\mu}$, a function of $\bm{\theta}$, represents a prediction of deterministic contributions to pulse arrival times, which includes our models, as well as deterministic noise transients such as exponential dips~\citep{GoncharovReardon2021}, and $\mathbf{C}$ built following established PTA practice (including analytic marginalization over the linearized timing model) \citep{vanHaasterenLevin2009,ArzoumanianBrazier2016,AntoniadisArzoumanian2022}. 
Time correlated noise for each pulsar is modeled with a Fourier basis, with the number of frequencies per pulsar set from our single-pulsar noise analyses \citep{EPTA_DR2_NOISE,PPTA_DR3_NOISE}. 
We also include a common stochastic gravitational wave background component represented by 30 Fourier frequencies. 
Model selection uses the Bayesian evidence $\mathcal{Z}$; Bayes factors $\mathcal{B}$ are ratios of evidences between competing models.

\subsection{Signal Propagation and Projection in Pulsar Timing Arrays}
\label{sec:method:projection}

A passing gravitational wave (GW) induces a fractional pulsar frequency shift
\begin{equation}
\frac{\delta\nu(t)}{\nu}
= F_{+,\times}(\theta,\phi,\psi)\,\Big[h_{+,\times}(t) - h_{+,\times}\!\big(t-\vec{n}\!\cdot\!\vec{r}\big)\Big],
\end{equation}
where $F_{+,\times}(\theta,\phi,\psi)$ is the geometric (quadrupolar) antenna response for a given pulsar to a GW source at a sky location $(\theta,\phi)$ relative to the pulsar's line of sight, and $\psi$ is the GW polarization angle. The first term is the Earth term (common to all pulsars) and the second is the pulsar term \citep{vanHaasterenLevin2010}. 

The projected timing residuals $\delta t (t)$ in each pulsar can be expressed in a compact format as: 
\begin{align}
    \delta t (t) &= \begin{bmatrix}F_{+} & F_{\times}\end{bmatrix}\begin{bmatrix}\cos2\psi & -\sin2\psi\\
\sin2\psi & \cos2\psi
\end{bmatrix}\begin{bmatrix}\delta t_{+}(t)\\
\delta t_{\times}(t)
\end{bmatrix}\,.
\label{eq:r_matrix}
\end{align}
where $\delta t_{+\times} (t) =   \int_{t_0}^t \, h_{+,\times} (t')\, dt'$ as the residuals can be obtained by integrating the GW strain. For full derivations and definitions, see Section B of our methods paper \citep{TomsonGoncharov2025}.

PTAs measure the \emph{post-fit} response obtained after a simultaneous fit of the pulsar timing model (spin and spindown). Denoting the pre-fit Earth-term response by $\delta t(t)$, the post-fit signal is
\begin{equation}
\delta t_{\rm post}(t)=\delta t(t)-\mathcal{P}_{\rm poly}\big[\delta t_{\rm E}\big](t),
\label{eq:postfit}
\end{equation}
where $\mathcal{P}_{\rm poly}$ is the least-squares projection onto $\{1,t,t^2\}$ over the data span. In what follows we apply this common projection to each signal model.

In this work we consider two signals: (i) a generic burst-with-displacement memory model, and (ii) a complete supermassive black hole binary (SMBHB) merger waveform including null memory. Because PTA pulsars lie at $\sim$kpc distances, the pulsar term yields at most a single jump (step) in the gravitational wave strain within a $\sim$decade data span, For robust detection we focus exclusively on the Earth term response from the signal models.

\subsection{The signal model for bursts with displacement memory}
\label{sec:method:burst}

In the burst (instantaneous) limit, the strain time series can be approximated as a step function in the metric traveling through space~\citep{vanHaasterenLevin2010} and is linearly polarized. In a suitable polarization basis (a convention), the memory can be written entirely in the \(+\) mode,
\begin{equation}
h_{+}(\vec{r}, t)=h_0\times \Theta\left[(t-t_0)-\vec{n}\cdot \vec{r}\right],
\label{eq:dcwave}
\end{equation}
where $h_0$ is the amplitude of the step, $t_0$ is the time at which the burst passes the Solar System Barycenter (SSB), $\vec{r}$ is the location of the pulsar relative to the SSB and $\vec{n}$ is the unit vector in the direction of the gravitational wave propagation and $h_{\times}(\vec{r}, t) = 0$. In~\citet{PshirkovBaskaran2010}, the amplitude of the step for SMBHBs is shown to be $h_0 \propto \mu/r$, where $\mu$ is the reduced mass of the binary and $r$ is the comoving distance.
See also Equation~4.5 in a more recent study by~\citet{ElhashashNichols2025}. 

The Earth-term strain induced by a gravitational-wave memory burst, as observed in pulsar timing residuals, can be expressed as 
\begin{equation}
\label{eq:burst_strain_projected}
h(t) = F(\theta, \phi, \psi) \cdot h_0 \cdot \Theta(t - t_0).
\end{equation}

In pulsar timing analyses, however, the measured quantity is the timing residuals, which is the time integral of $h(t)$. An instantaneous step in the strain therefore produces a linearly growing (ramp) signature in the residuals at times after the burst.

\subsection{SMBHB mergers with null memory: a complete model}
\label{sec:method:nr}

The burst-with-memory approximation is source-agnostic. It models a displacement without assuming a specific origin. 
As such, it can capture signals produced by supermassive black hole binaries as well as by more exotic mechanisms (e.g., cosmic-string cusps ~\citep{JenkinsSakellariadou2021}), provided the memory growth is fast compared to the PTA cadence. However, it does not encode the accompanying oscillatory waveform, nor does it cover other memory effects (e.g., spin or center-of-mass memory). For sources with slower memory buildup or distinctive morphology, dedicated signal models are preferable.

While this enables a broad, \emph{source- and morphology-agnostic} search, it neglects the physical context and waveform features expected from realistic sources.

In this section, we refine our methodology by adopting a physically motivated model for mergers of SMBHBs, which includes the contribution of the null memory. 
There are several key advantages to using the full signal model for null memory searches:
\begin{enumerate} 
    \item \textit{Physics-motivated}: SMBHBs are the primary expected source of memory in PTAs. The full waveform captures the complete time-domain structure of the gravitational wave signal, including the inspiral, merger, ringdown, and the gradual buildup of null memory. This offers a dedicated representation of the waveform from SMBHB mergers.
    \item \textit{Robustness to noise}: Generic burst models, while flexible, are more prone to capturing unmodeled noise artifacts. 
    \item \textit{Astrophysical Interpretability}: The full waveform model is parametrized by the chirp mass $\mathcal{M}$, luminosity distance $D_\text{L}$, mass ratio $q$ and sky location $(\theta,\phi)$ of the SMBHB. This allows for direct inference of source properties from data such as the expected rates and number density of SMBHB mergers. This may offer insights into the assembly history of SMBHB and galaxy evolution.

   \item \textit{Observable post-fit amplitude for SMBHBs}: The burst template assumes an instantaneous step at merger, whereas the SMBHB memory builds up gradually during the late inspiral~\citep{Favata2009} and is captured by our SMBHB-merger waveform. After the timing-model fit ( pulsar spin and spindown), part of this slow rise is absorbed, at the merger epoch, the burst (step) template produces a larger post-fit bump than the SMBHB waveform, i.e., it overestimates the observable (post-fit) amplitude. This effect is quantified below and in our companion methods paper~\citep{TomsonGoncharov2025}.
\end{enumerate}

We use the \texttt{NRHybSur3dq8\_CCE} surrogate waveform model developed by ~\citet{YooMitman2023} which combines numerical relativity, post-Newtonian, and effective-one-body waveforms through hybridization. This model includes both the oscillatory inspiral-merger-ringdown (IMR) signal and the null memory contribution, capturing the full gravitational-wave strain with high fidelity. The strain polarizations $h_+ (t)$ and $h_\times (t)$ are computed from the dominant radiative modes, including memory in the non-oscillatory $(2,0)$ mode. The timing residuals are then obtained using Equation~\eqref{eq:r_matrix}. The waveform is parameterized by physical quantities such as chirp mass, mass ratio, spin, luminosity distance, inclination and sky location of the binary, enabling direct astrophysical inference from PTA signals.

\begin{figure}[htbp]
  \centering
  \includegraphics[width=0.46\textwidth]{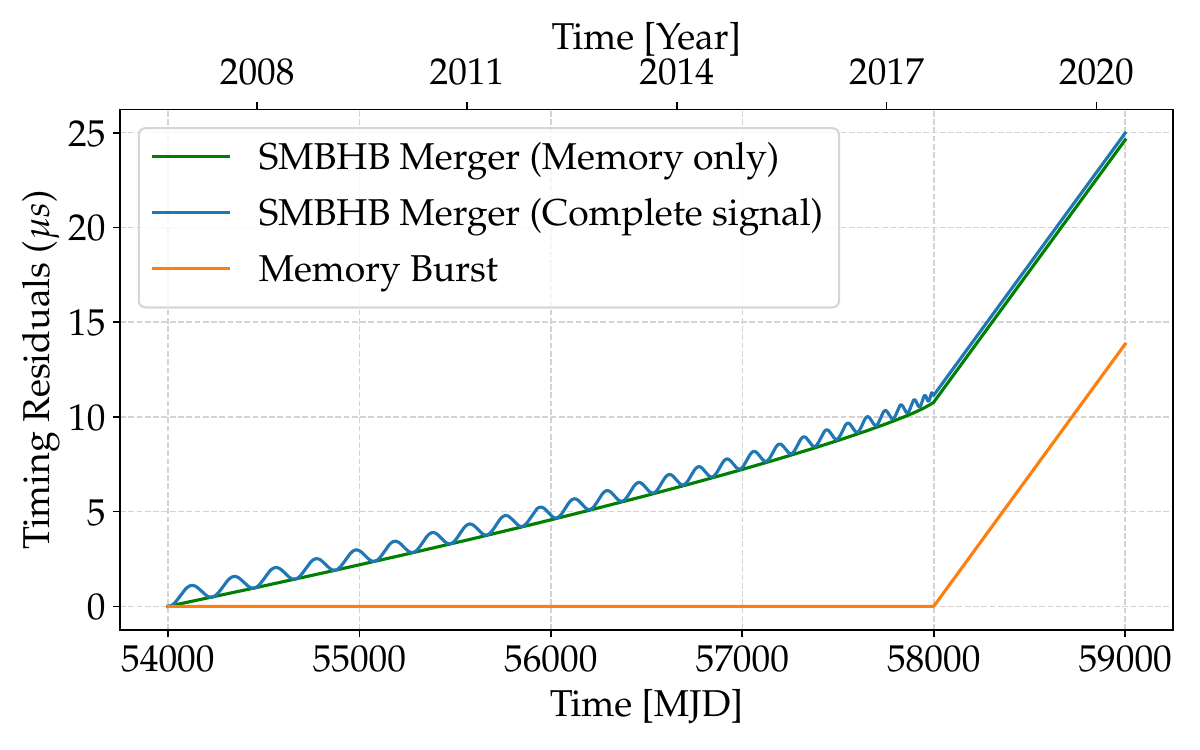}
  \caption{
  PTA timing residuals from a merger of a non-spinning supermassive black hole binary (SMBHB) with parameters
  $\mathcal{M} = 10^{10} M_{\odot}$, $q=1$, and $D_\text{L} = 100$ Mpc. observed by a pulsar at $(\text{ra},\text{dec}) = (258.4564^{\circ},7.7937^{\circ})$. The merger occurs at $\mathrm{MJD}=58000$. Green : SMBHB merger model with only memory, Blue : SMBHB merger model (complete signal including IMR+memory),  Orange : Memory burst model. All curves are normalized to the same final memory offset.}
  \label{fig:burst_vs_merger}
\end{figure}

Figure~\ref{fig:burst_vs_merger} shows the pre-fit PTA responses on a pulsar located at a right ascension and declination of $(\text{ra},\text{dec}) = (258.4564^{\circ},7.7937^{\circ})$ for a fiducial equal-mass source ($\mathcal{M}=10^{10}\,M_\odot$, $D_L=100$\,Mpc, $(\text{ra},\text{dec}) = (0^{\circ},0^{\circ})$) using both the memory-burst (ramp) template and our SMBHB merger model.
However, PTAs are not sensitive to this time series in full. 
After subtracting the timing model (spin and spindown), part of the SMBHB signal's slow, pre-merger rise is absorbed by the fit, so the observable post-fit bump is smaller than for a ramp with the same final memory offset. Quantitatively, the noise-agnostic RMS difference between the post-fit residuals of the two templates is $0.399\,\mu\mathrm{s}$ when their final memory offsets are matched. If the burst amplitude is allowed to float to minimize this mismatch, the RMS drops to $0.26\,\mu\mathrm{s}$, but the best-fit burst amplitude is $11.8\%$ smaller than the merger model’s offset (and the corresponding biases in $\mathcal{M}$ or $D_L$). The discrepancy shrinks by $\sim0.1\,\mu\mathrm{s}$ when the mass ratio changes from $q=1$ to $q=7$. These results motivate the physically consistent IMR+memory waveform to avoid inflated signals and parameter estimation biases. The corresponding post-fit residual panels and extended comparisons are shown in Fig.~4 of~\citet{TomsonGoncharov2025}.

\subsection{Factorized posterior approximation}
\label{sec:method:approx}

The Factorized Posterior (FP) approach is an accelerated Bayesian approach that constructs the full-PTA posterior from single-pulsar posteriors, provided the full posteriors can be represented as a product of individual posteriors. 
This is the case when neglecting inter-pulsar correlations for stochastic time-correlated signals.
In this Section, we review the application of FP to PTA data analyses. 
First, we explain the existing methodology by~\citet{SunBaker2023}.
Second, we introduce the extensions of this methodology to allow for smooth and continuous posterior approximation, which are made possible by Kernel Density Estimation (KDE) and normalizing flows~\citep[flows,][]{SrinivasanCrisostomi2024}.

Following~\citet{SunBaker2023}, the intrinsic (projected per-pulsar) amplitude of memory $h_i$ for the $i$-th pulsar is related to the global (extrinsic) memory amplitude offset $h_{0}$ via the geometric projection factor as $h_i = h_0 F_i(\theta, \phi, \psi)$, where \(F_i\) is the geometric projection defined in Sec.~\ref{sec:method:burst} (Eq.~\eqref{eq:burst_strain_projected}). This factor determines the sign and magnitude of the intrinsic per-pulsar amplitude, indicating whether the timing residuals are advanced or delayed by the burst.

Once the intrinsic amplitudes $h_i$ are computed, the FP approach combines the pulsar-term likelihoods 
\begin{equation}
\label{eq:FP}
    p(\delta t | h_0, t_0, \theta, \phi, \psi)  = \prod_{i=1}^N p_i (\delta t | h_i, t_0)
\end{equation}

where $t_0$ is the burst epoch and N is the number of pulsars. Unlike the original implementation of~\citet{SunBaker2023}, which relies on precomputed lookup tables over a grid of projected amplitudes, our approach uses smooth approximations of the single-pulsar posteriors using Kernel Density Estimation (KDE) and normalizing flows. 

KDE is a non-parametric way of approximating a set of samples into a continuous (analytical) probability density function \citep{Silverman1986}. Using KDEs within a factorized posterior framework for PTA analyses has been discussed previously \citep{LambTaylor2023}.
It defines a kernel at every point in the parameter space and assigns a density to each point. We construct bounded kernel density estimates (KDEs) from MCMC samples over the 2D space of intrinsic per-pulsar amplitude $h_i$ and burst epoch $t_0$. While we start with a Gaussian kernel and use the ``Scott'' bandwidth selection rule as a baseline \citep{Scott2015}, in practice, we fine-tune the kernel width heuristically based on the shape and structure of the posterior to ensure adequate resolution and smoothness. We implement reflective padding to handle boundary effects that can lead to underestimated densities near the edges of the prior volume. This allows the KDE to correctly capture the probability mass near the limits without artificially truncating or distorting the distribution. As the memory signal can appear in pulsars with either positive or negative amplitude depending to sky geometry, we partition the MCMC samples into positive and negative amplitude subsets and build separate KDEs for each. The sign of the extrinsic amplitude $h_0$ during global inference selects which KDE to use for each pulsar's contribution. While KDEs provide a reasonable approximation of the 2D posteriors, they are computationally expensive during evaluation, since each likelihood call involves summing over all single pulsar MCMC samples. This computational cost motivates the adoption of faster alternatives like normalizing flows.

Normalizing flows are a more efficient approach based on machine learning to approximate a complex \emph{target} per–pulsar posterior \(p(x)\) (with parameters \(x\), e.g., \(x=(h_i,t_0)\)) with a flexible \emph{model} density \(p_\theta(x)\). The model is built by transforming a simple \emph{base} distribution \(p_Z(z)\) (typically \(\mathcal N(0,I)\)) through an invertible map \(f_\theta:\mathbb R^d\!\to\!\mathbb R^d\) (a normalizing flow) \citep{JimenezRezendeMohamed2015}. 
Writing \(x=f_\theta(z)\) and \(z=f_\theta^{-1}(x)\), the change-of-variables formula gives the density
\begin{equation}
p_\theta(x)=p_Z\!\big(f_\theta^{-1}(x)\big)\,\left|\det \frac{\partial f_\theta^{-1}(x)}{\partial x}\right|
\end{equation}

We fit \(\theta\) by maximizing the likelihood (equivalently, minimizing the negative log-likelihood) on samples from \(p(x)\).
After training, the sampling is fast (draw \(z\sim p_Z\) , map \(x=f_\theta(z)\)) and density evaluation uses \(z=f_\theta^{-1}(x)\) and the Jacobian.
In this work we adopt Masked Autoregressive Flows (MAFs) \citep{PapamakariosNalisnick2019}, a widely used architecture for density estimation, and adopt the \texttt{floZ} package from \citet{SrinivasanCrisostomi2024}. In our factorized-posterior pipeline, flows are much faster than KDE factors because evaluation and sampling do not scale with the number of stored samples.

To ensure robust modeling of bounded parameters and to preserve the posterior structure, we apply a parameter-wise logit transformation to the posterior samples before training. This maps each parameter from its prior box $[a,b]$ to the whole real line, eliminating hard cutoffs that otherwise cause discontinuities and spurious density outside the prior support.

Let \(r=(x-a)/(b-a)\in(0,1)\). We define
\begin{equation}
y=\operatorname{logit}(r)=\log\frac{r}{1-r} = \log\!\frac{x-a}{\,b-x\,}
\end{equation}
and $x=a+(b-a)\,\sigma(y)$, where \(\sigma(y)=1/(1+e^{-y})\) is the logistic function (Equation 1.10 of \citet{GelmanCarlin2014}).
The 1D Jacobian is 
\begin{equation}
\left|\frac{dy}{dx}\right|=\frac{1}{(b-a)}\cdot\frac{1}{r(1-r)}=\frac{b-a}{(x-a)(b-x)} 
\end{equation}
Because this transform is applied component-wise, the Jacobian is diagonal in n dimensions (n parameters) and the total log-Jacobian is the sum over components.

After the logit step, the samples are whitened via eigenvalue decomposition \citep{HeavensFantaye2017}. 
Namely, the samples are projected along the eigenvectors of the covariance matrix of the samples, and rescaled by the square root of the eigenvalues. This ensures the flow operates in a de-correlated and standardized latent space, facilitating efficient training.

The normalizing flow is trained on the whitened logit–space samples $\tilde y$. During training, the log-Jacobian of this fixed preprocessing does not depend on the flow parameters and therefore does not affect optimization. At evaluation time, however, when reporting densities back in the original variables $x$, we apply the change-of-variables rule to account for all invertible steps:
\begin{equation}
\label{eq:nf_covar_total}
\log p_\theta(x)
= \log p_\theta(\tilde y)
+ \log\!\left|\det\frac{\partial \tilde y}{\partial y}\right|
+ \log\!\left|\det\frac{\partial y}{\partial x}\right|
\end{equation}
Here $\log p_\theta(\tilde y)$ is provided by the trained flow, $\displaystyle \frac{\partial \tilde y}{\partial y}$ and  $\displaystyle \frac{dy}{dx}$ are the Jacobians. To generate samples in the original parameter space after training, we invert the pipeline by first drawing the samples $\tilde y$ from the flow, applying inverse whitening and inverse logit transformations respectively.

Training the flow model involves tuning key parameters like the number of flow layers, hidden units, learning rate, and batch size, all of which impact performance and convergence. After model construction, we generate samples from the model in the original parameter space by the inverse transformation.

\label{sec:method:flybys}

\section{Results}
\label{sec:results}

\subsection{Results for generic memory bursts}
\label{sec:results:burst}

We perform a search for generic memory bursts, defined as per the methodology of Section~\ref{sec:method:burst}. 
We perform a simultaneous fit of the signal model -- which includes our signal (memory bursts), common-spectrum noise process attributed to the gravitational wave background, as well as the noise model -- to the data. 
We find {$\ln \mathcal{B} = 0.712$} for the signal component in 10-year EPTA DR2, $\ln \mathcal{B} = -0.309$ for 25-year EPTA DR2, and $\ln \mathcal{B} = 0.432$ for PPTA DR3.

For computational efficiency, we modeled the gravitational wave background (GWB) as a common-spectrum red noise process without Hellings-Downs (HD) spatial correlations in our main analysis. However, for the 10-year EPTA DR2 dataset, this approximation led to an outlier with a higher Bayes factor. To address this, we include the HD-correlated GWB model for this particular dataset, which yielded the Bayes factor we reported above. We find no evidence of gravitational wave memory bursts in these data.  With no evidence for memory bursts, based on the results of parameter estimation\footnote{We show and discuss more broadly the results of parameter estimation for all parameters of the memory burst model in Appendix~\ref{sec:appendix:pe}.}, we compute upper limits. 

\begin{figure}[htbp]
  \centering
  \includegraphics[width=0.48\textwidth]{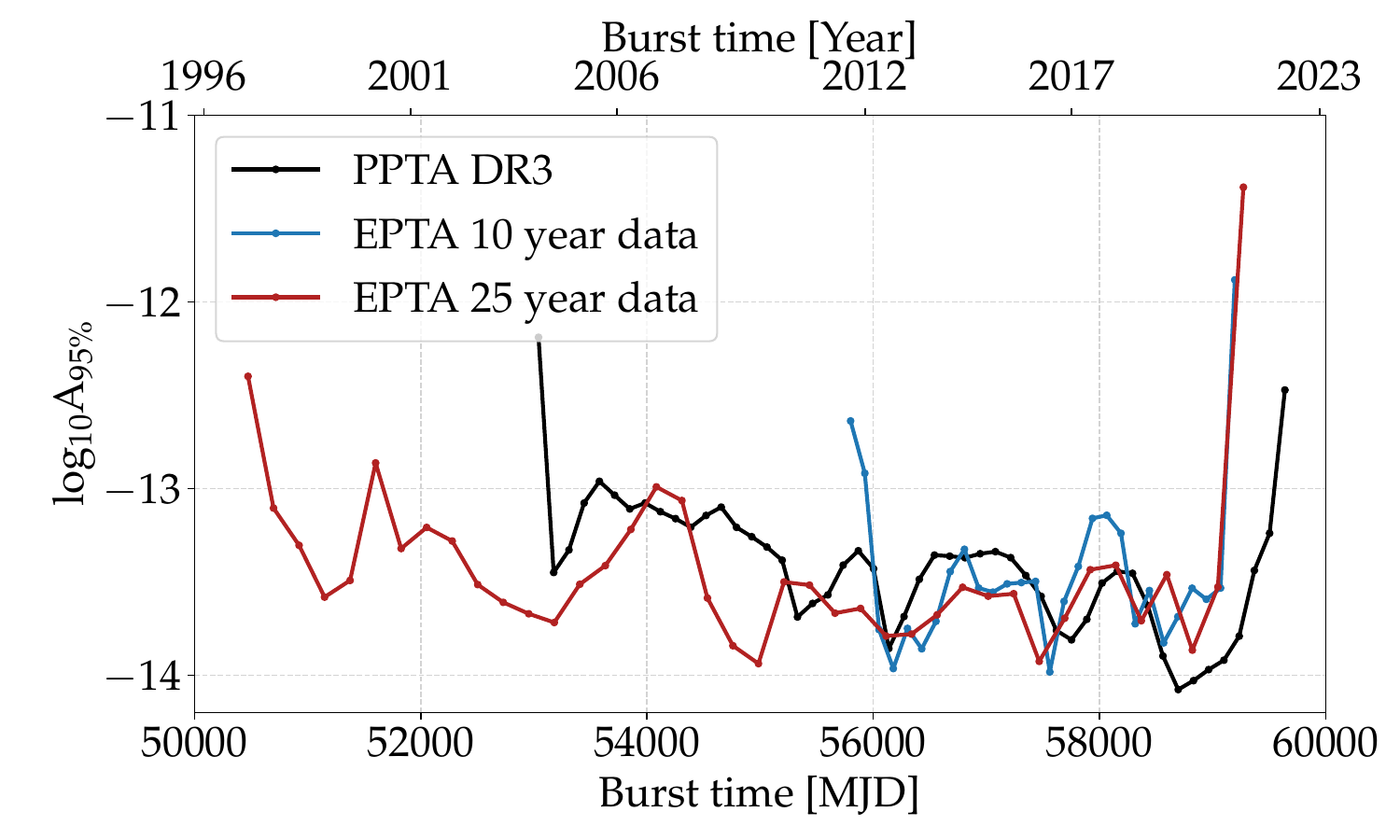}  
  \caption{
  Upper limits on strain amplitude $h_0$ of generic displacement memory bursts as a function of burst time at 95\% credibility. Other burst parameters are marginalized over. }
  \label{fig:burst_mjd}
\end{figure}

We follow the procedures outlined in~\citet{NG_15_MEM} to compute upper limits in two ways:
(1) as a function of burst epoch, marginalizing over sky location and polarization; and
(2) as a function of sky position, marginalizing over burst time and polarization.
In Figure~\ref{fig:burst_mjd}, we show $95\%$ upper limits on the memory burst strain amplitude as a function of burst time. These limits vary due to the inhomogeneous observational coverage across different years.
The 25-year EPTA DR2 extends to year $1996$, being our only source of information about memory bursts between years $1996$ and $2004$. 
The PPTA has legacy observations dating back to $1994$, but those epochs are not included in PPTA DR3 and are not used in this work.
At about year $2019$, we achieve the most constraining limits with the PPTA data, ruling out strain amplitudes of $h > 10^{-14}$ between MJD 58600 and 58800. 
The sensitivity of the PPTA has been enhanced during this period due to the installation of the ultra-wideband receiver.
It is visible that our sensitivity to memory bursts degrades towards the edges of the observation span. 
Memory bursts, which would have occurred beyond the observation span, would introduce pulsar timing delays that grow mostly linearly with time. 
Currently, PTAs can not distinguish these signatures from errors in pulsar spin frequencies. 
Towards the middle of the observation span, the limits vary by up to one order of magnitude due to the irregularity of observations.

\begin{figure*}[htbp]
  \centering
  \gridline{
    \fig{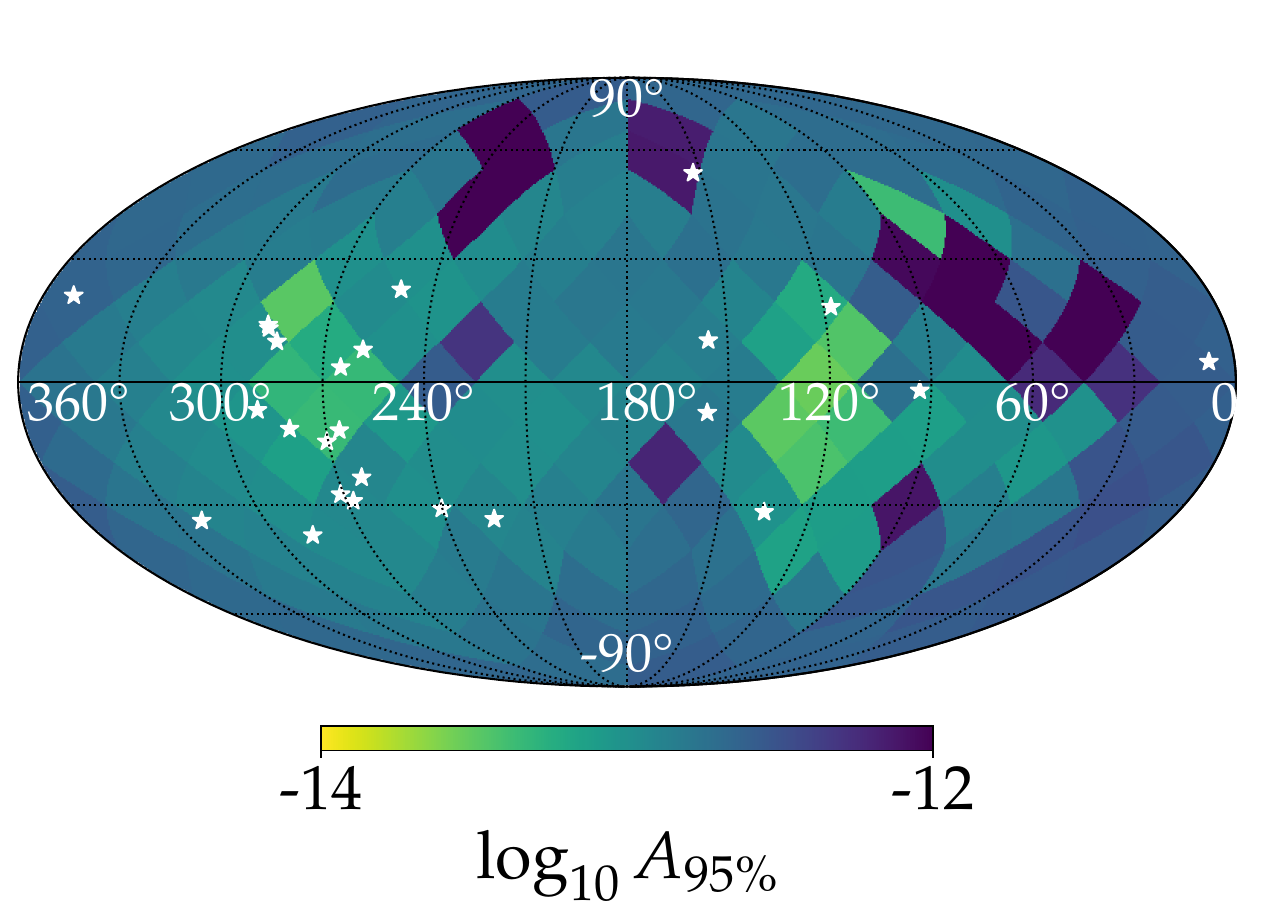}{0.32\textwidth}{\label{fig:skymap:epta10}}
    \fig{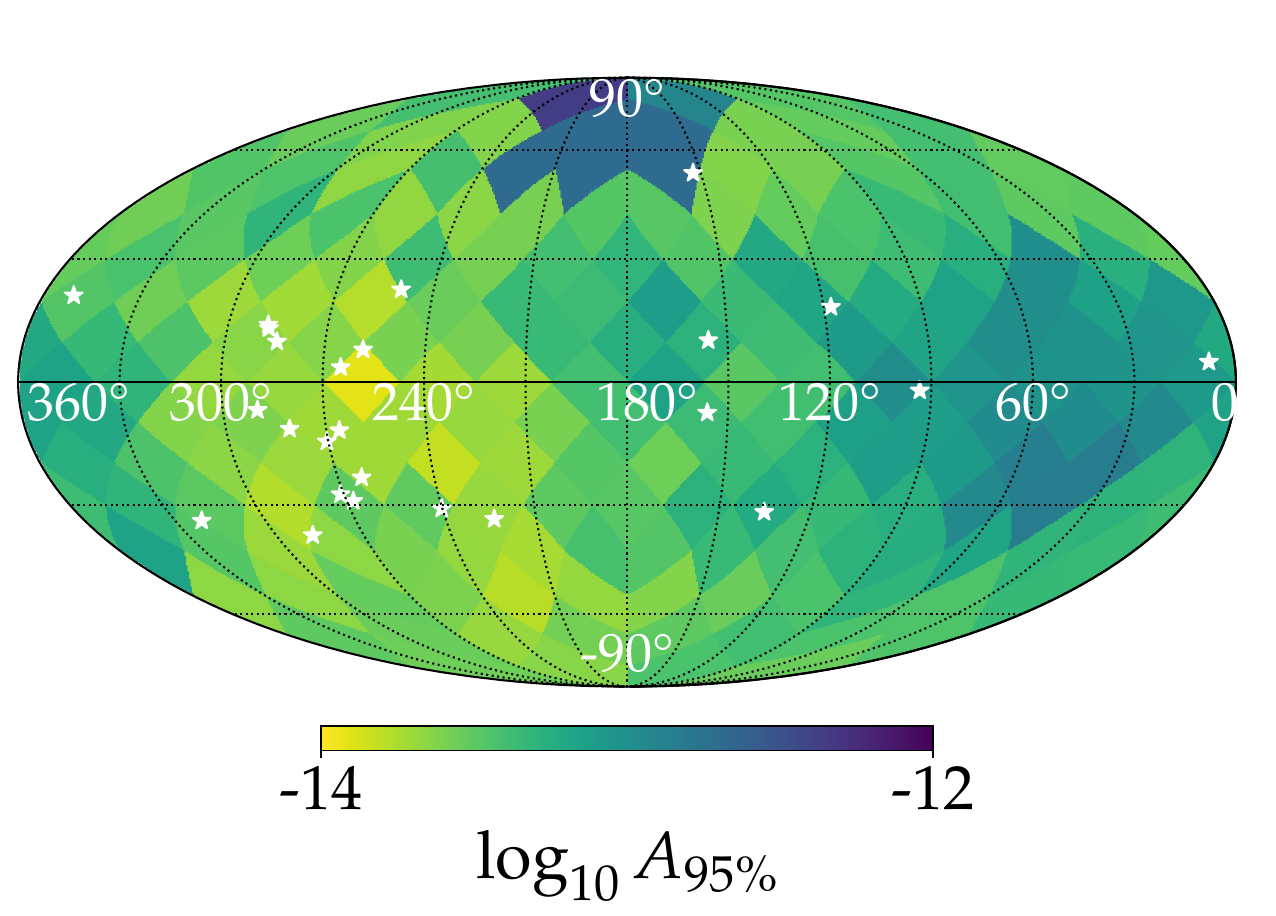}{0.32\textwidth}{\label{fig:skymap:epta15}}
    \fig{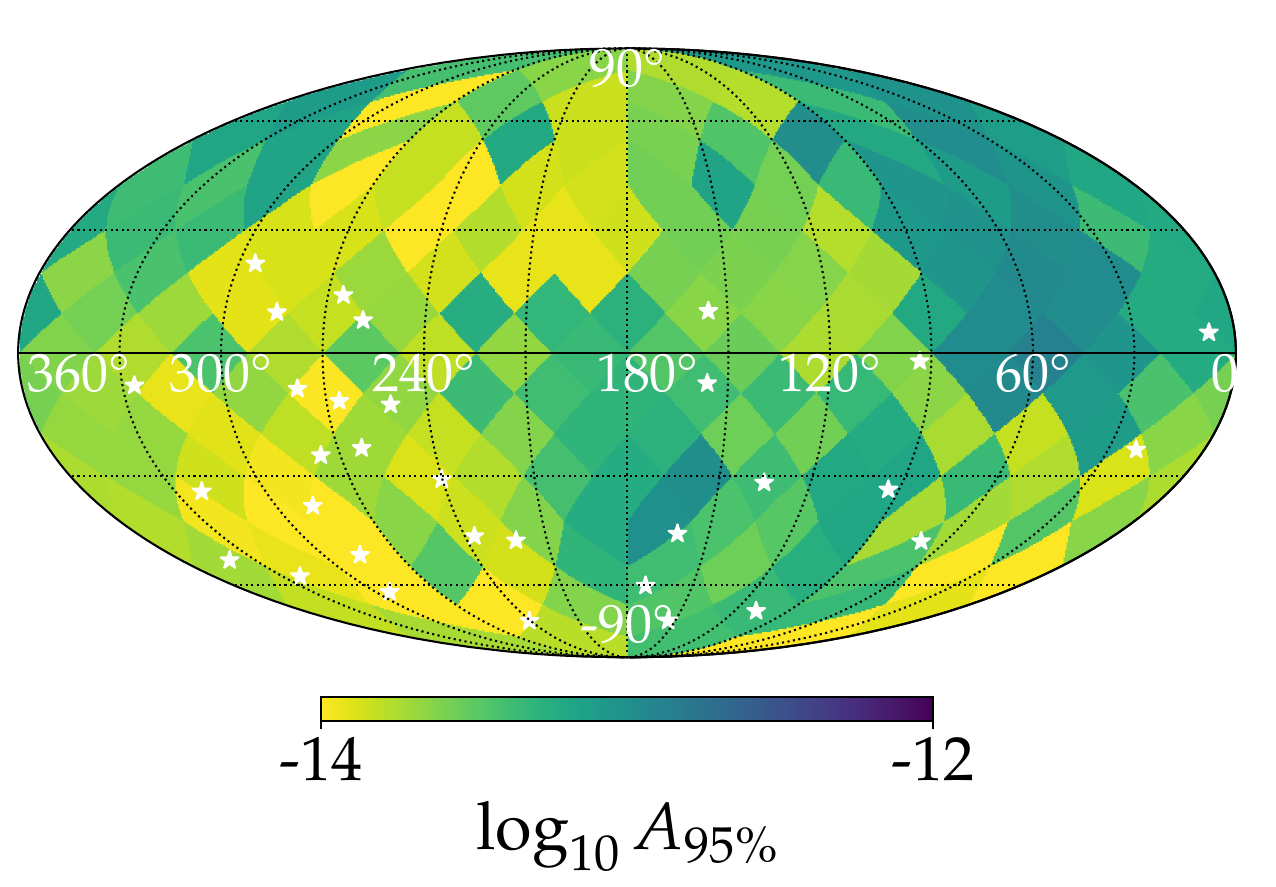}{0.32\textwidth}{\label{fig:skymap:ppta}}
  }

  \caption{Upper limits on the strain amplitude $h_0$ of generic displacement-memory bursts, shown in colour, as a function of sky position $(\theta,\phi)$ at 95\% credibility (other burst parameters marginalised over).  
  The three panels correspond to the \textit{EPTA 10‐year}, \textit{EPTA 25‐year}, and \textit{PPTA DR3} data sets, respectively (left → right). White stars mark the pulsar positions for each PTA.}
  \label{fig:skymap}
\end{figure*}

In Figure~\ref{fig:skymap}, we show upper limits on the memory burst strain amplitude as a function of the sky position. 
We place limits on memory strain across $192$ sky pixels of equal area. 
The burst epoch and polarization of the GW source are marginalized over.
Due to the non-uniform distribution of pulsars in our PTA, sensitivity to gravitational wave bursts varies significantly across the sky. 

Thus, we perform a separate calculation of an upper limit for every sky location.

The most constraining limits for PPTA DR3 are obtained at $(\theta,\phi)=(292.5^{\circ},-66.444^{\circ})$, and the least constraining limits are obtained at $(\theta,\phi)=(56.25^{\circ},0^{\circ})$. 
The most constraining limits for 10-year EPTA DR2 are obtained at $(\theta,\phi)=(281.25^{\circ},19.471^{\circ})$, and the least constraining limits are obtained at $(\theta,\phi)=(33.75^{\circ},19.471^{\circ})$. 
The most constraining limits for 25-year EPTA DR2 are obtained at $(\theta,\phi)=(285^{\circ},-54.341^{\circ})$, and the least constraining limits are obtained at $(\theta,\phi)=(78.75^{\circ},-19.471^{\circ})$. The spatial variation in the upper limits reflects the distribution and sensitivity of pulsars across the sky in each array. As expected, the most constrained upper limits are obtained in regions where the pulsar sky coverage is densest. 
For instance, the lowest upper limits in all three datasets—PPTA DR3, 10-year EPTA DR2, and 25-year EPTA DR2—are found in sky regions that have more pulsars. 
Notably, the most-constrained directions for the 10-year and 25-year EPTA datasets are broadly consistent, suggesting that extended timing baselines improve sensitivity without significantly shifting the sky region of maximum constraint. 
In contrast, the least constraining limits tend to occur in regions with sparse or no nearby pulsars, demonstrating the inherent anisotropy in PTA sensitivity. The upper limits as a function of burst sky location from the 25-year EPTA DR2 dataset are lower than the limits from the 10-year dataset. Although some of the early EPTA data are noisier, the post-fit memory response gains sensitivity from a longer observation time.
This is also consistent with other long-timescale signal searches (e.g., ultralight dark matter), where extending the observation baseline tightens constraints despite mixed data quality \citep{SmarraGoncharov2023}.

\subsection{Search for SMBHB mergers with null memory}
\label{sec:results:nr}

We perform a search for SMBHB mergers in PPTA DR3 and 25-year EPTA DR2 using the \nolinkurl{NRHybSur3dq8_CCE} waveform, which includes strain time series for the inspiral and merger of SMBHBs, the ringdown of the remnant, and the null displacement memory. 
We find $\ln \mathcal{B} = -0.081$ for 10-year EPTA DR2 $\ln \mathcal{B} = -0.769$ for 25-year EPTA DR2, and $\ln \mathcal{B} = -0.679$ for PPTA DR3. 
Therefore, we find no evidence of a signal. Table ~\ref{tab:bf} summarizes the Bayes factor values obtained from searches on different datasets using the models used in the paper.

\begin{figure*}[htbp]
  \centering
  \gridline{
    \fig{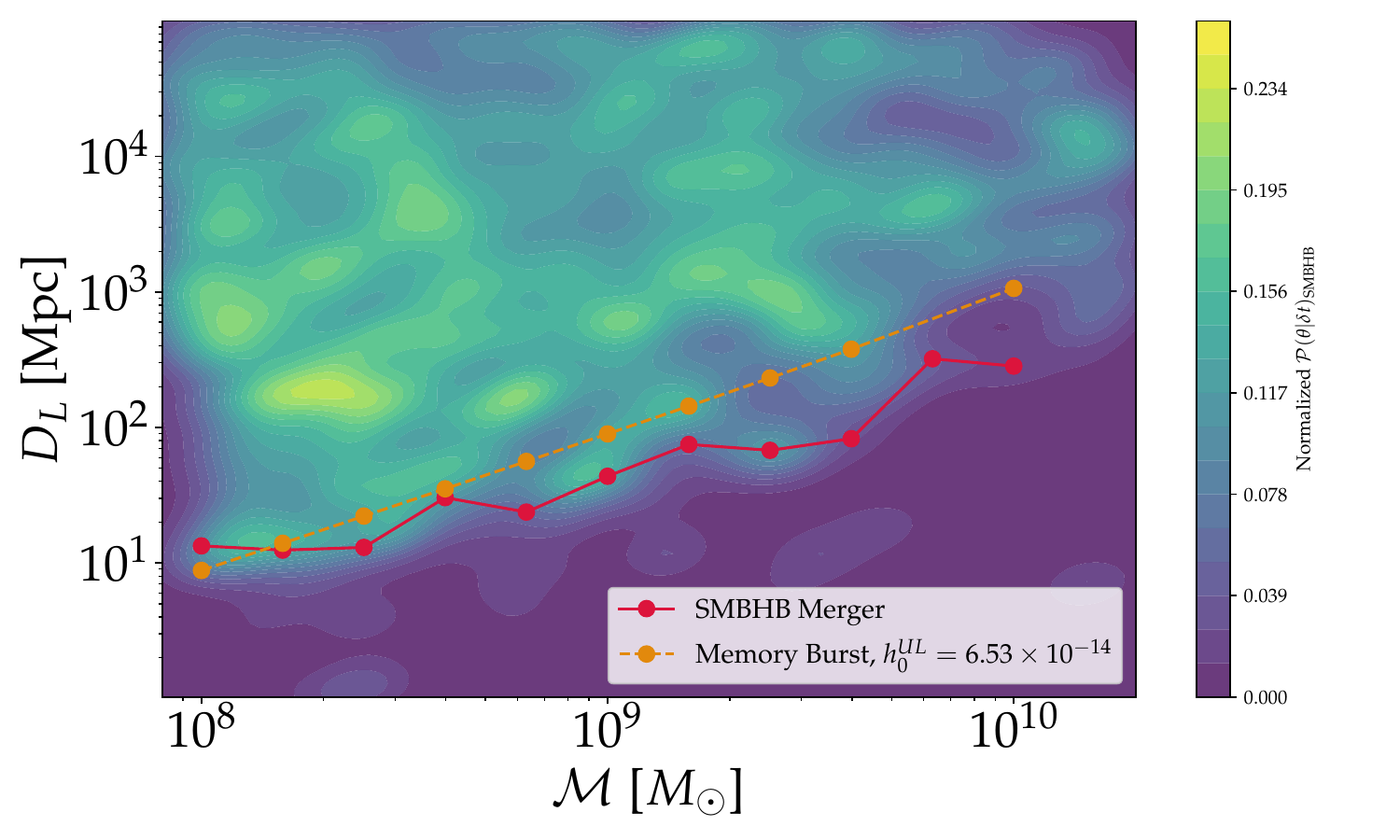}{0.49\textwidth}{(a) 25-year EPTA DR2\label{fig:Dist_LL:epta25}}
    \fig{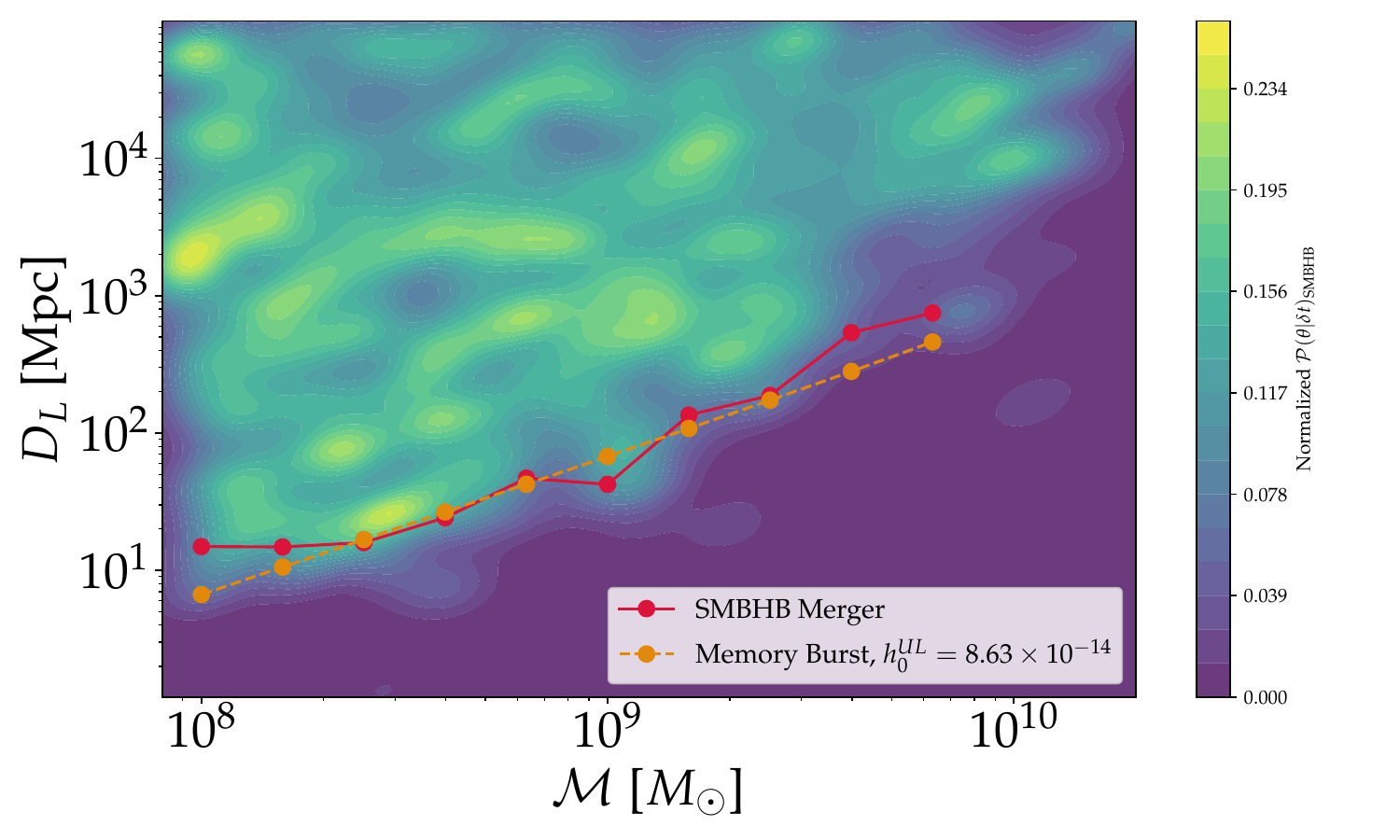}{0.49\textwidth}{(b) PPTA DR3\label{fig:Dist_LL:ppta}}
  }

  \caption{Lower limits on the luminosity distance $D_\text{L}$ as a function of source-frame chirp mass $\mathcal{M}$. \textbf{This comparison is valid for equal mass binaries.}}
  \label{fig:Dist_LL}
\end{figure*}

We present our constraints on chirp masses $\mathcal{M}$ and luminosity distances $D_\text{L}$ of merging SMBHBs in Figure~\ref{fig:Dist_LL}. 
We show both the posterior density and the lower limits at $95\%$ credibility on $D_\text{L}$ up to which an SMBHB with a given $\mathcal{M}$ can be detected. 
For comparison, we also show limits based on the estimation of $h_0$ of the memory burst model, assuming the burst strain amplitude corresponds to that of an SMBHB merger.  In Section~\ref{sec:method:burst} we mention the scaling $h_0 \propto \mu/r$~\citep{PshirkovBaskaran2010}, where $\mu$ is the reduced mass and $r$ is the comoving distance. 
For Figure~\ref{fig:Dist_LL}, we reparameterize this scaling in terms of source-frame chirp mass and luminosity distance. 
Using the definitions of $(\mu,r)$ and $(\mathcal{M},D_\text{L})$, and assuming the mass ratio $q = 1$, we can compute the chirp mass, $\mathcal{M} = (1+z)\mu ({1+q})^{4/5}/({q}^{2/5})$ and $D_\text{L} = (1+z) r$. 
Since the strain samples inferred from the search is in the observer frame, an additional $(1+z)$ factor is included to convert the chirp mass to the source frame. 
Based on the cosmological model from  \citet{PlanckCollaborationAghanim2020}, we convert the comoving distance $r$, to redshift $z$, and then to luminosity distance $D_\text{L}$. Using this mapping, we convert the upper limits on $h_0$ obtained from the burst search into lower limits on $D_\text{L}$ for each value of $\mathcal{M}$.

\begin{table}[ht]
  \setlength{\tabcolsep}{39pt}   
  \renewcommand{\arraystretch}{1.3}   
  \begin{tabular}{|c|c|}
    \hline
    \textbf{Dataset} &  \textbf{$\ln \mathcal{B}$} \\ \hline\hline
    \multicolumn{2}{|c|}{Memory burst model}  \\ \hline
    EPTA  10-year   & $0.712$  \\
    EPTA  25-year   & $-0.309$   \\ 
    PPTA  DR3   & $0.432$   \\ \hline
    \multicolumn{2}{|c|}{SMBHB merger model} \\ \hline
    EPTA  10-year   & $-0.081$   \\
    EPTA  25-year   & $-0.769$   \\
    PPTA  DR3   & $-0.679$   \\
    \hline
  \end{tabular}
  \caption{Bayes Factors.}
  \label{tab:bf}
\end{table}

\begin{figure*}[htbp]
  \centering
  \gridline{
    \fig{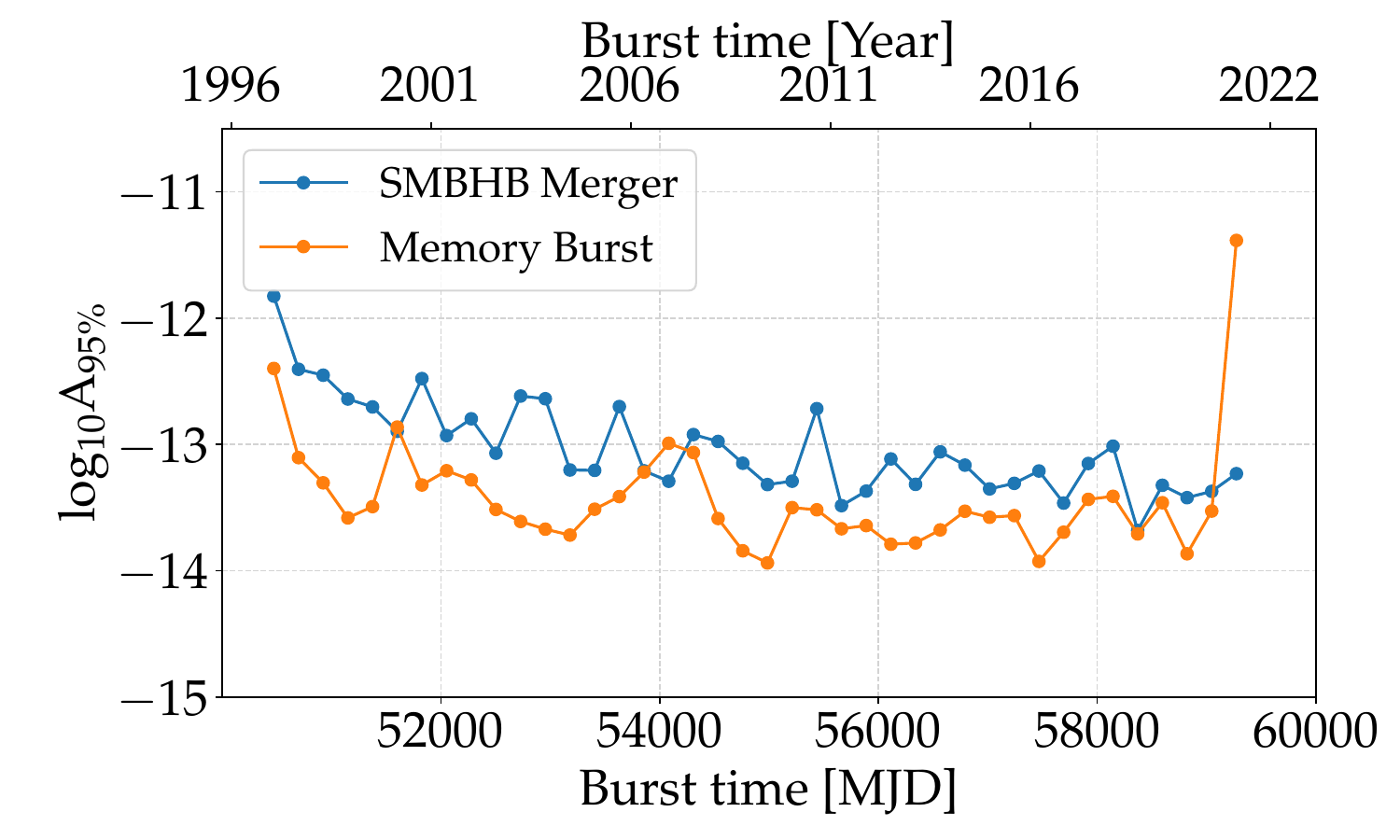}{0.49\textwidth}
        {(a) 25-year EPTA DR2\label{fig:nr_amp:epta25}}
    \fig{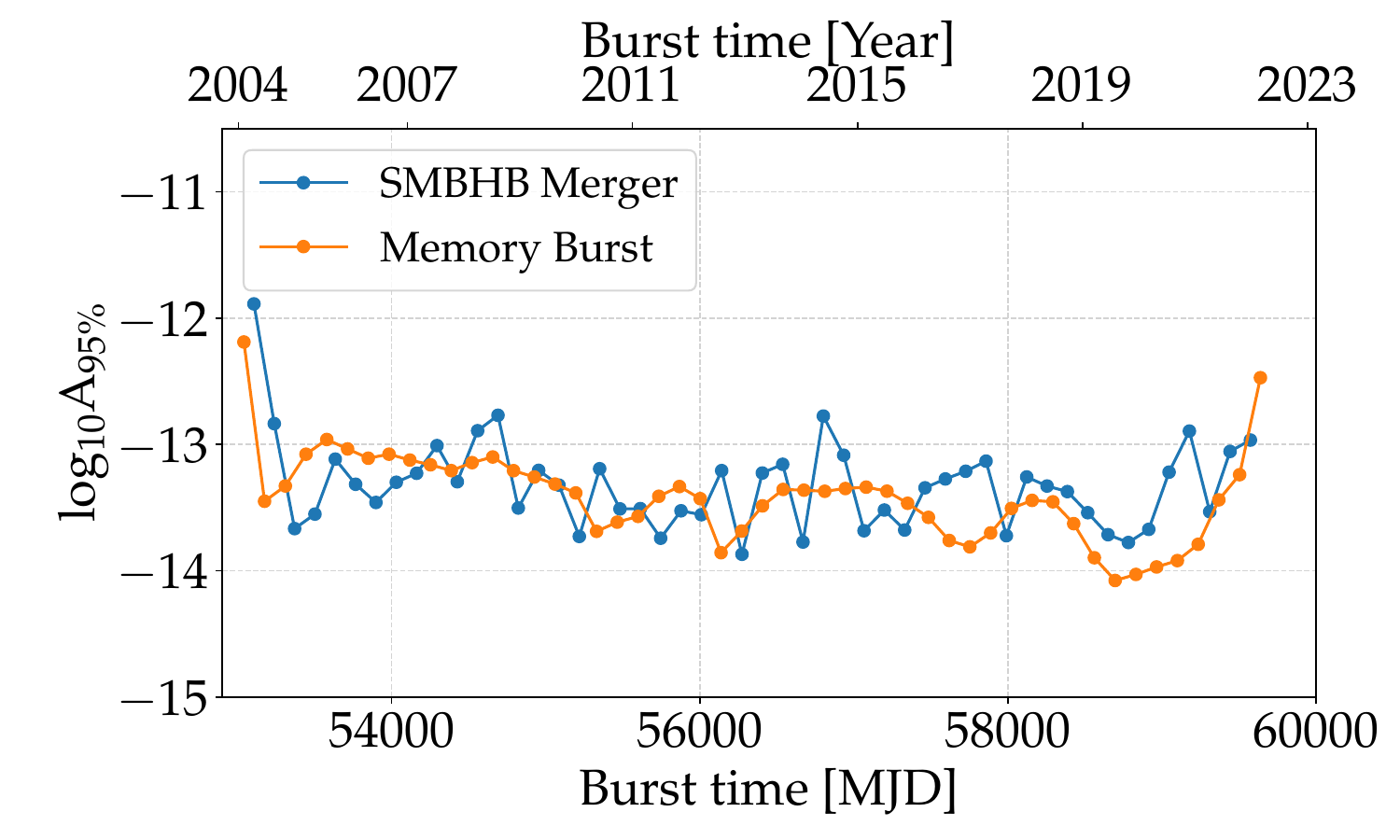}{0.49\textwidth}
        {(b) PPTA DR3\label{fig:nr_amp:ppta}}
  }

  \caption{Upper limits on the strain amplitude $h_{0}$ of generic memory bursts (orange) compared to the SMBHB merger model (blue) as a function of burst epoch. The merger model includes the gradual memory buildup during the inspiral, so the post-fit residuals are smaller than those of the burst model. As a result, the merger waveform yields slightly weaker upper limits compared to the generic burst model, but provides a more accurate representation of the expected memory signal. Unlike the burst model, the merger model limits do not degrade as sharply near the observation edges due to the contribution of the inspiral component.}
  \label{fig:nr_amp}
\end{figure*}

Using the same SMBHB memory scaling $h_0 \propto \mu/r$ and the mapping described above, we also convert the posterior samples in $(\mathcal{M}, D_\text{L}, q)$ obtained with the SMBHB-merger waveform model back into memory strain amplitudes $h_0$. From these samples we construct upper limits on the net null-memory strain amplitude as a function of the SMBHB merger (burst) time. The results are shown in Figure~\ref{fig:nr_amp}, where they are also compared to the results from the displacement-memory burst search in Figure~\ref{fig:burst_mjd}.

The limits for the NR model are slightly weaker (\textit{i.e.}, higher in strain and, correspondingly, lower in distance) than those from the generic burst model. This behavior arises because the burst model overestimates the observable residuals - it assumes an instant increase in the timing residuals, leading to artificially \emph{constraining} upper limits.
In contrast, the physically motivated NR waveform includes the gradual accumulation of memory during the inspiral. Consequently, the post-fit residuals for the NR model are smaller than for the burst model, leading to slightly higher upper limits on strain for the same dataset.
Unlike the limits for the memory burst model, the limits for the NR model do not degrade as quickly towards the edges of the observation time because of the contribution of the inspiral component of the signal.

\subsection{Factorized posterior comparison}

\begin{figure}[htbp]
  \centering
  \includegraphics[width=0.5\textwidth]{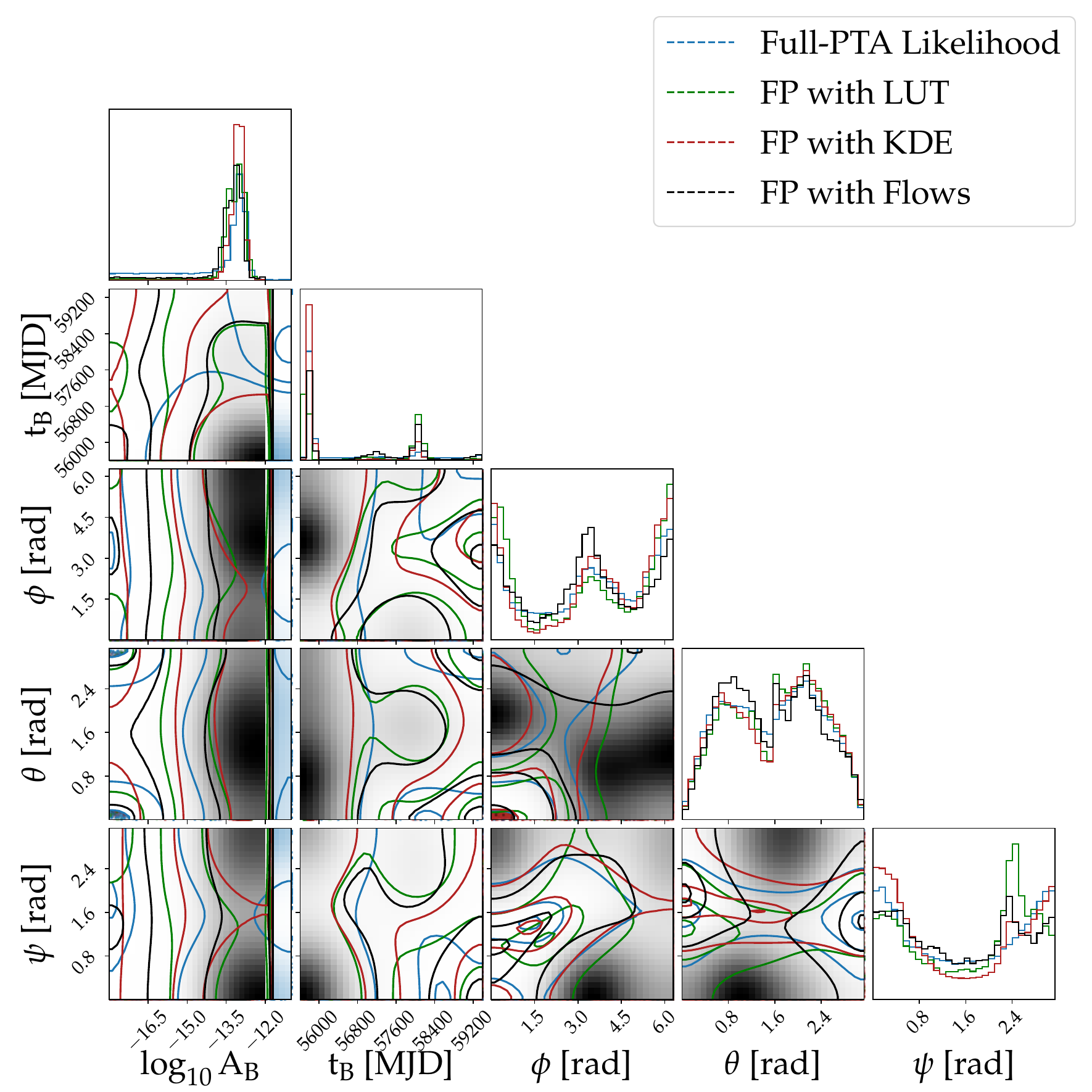}  
  \caption{
  Comparison of posterior distributions on the global memory burst parameters (amplitude, burst epoch, sky position, and polarization) using three inference methods applied to the 10-year EPTA dataset. The full-PTA likelihood results are shown in blue, FP with LUT in green, FP with KDE approximation in red, and FP with normalizing flows in black. 
  }
  \label{fig:dr2new_fp_comparison}
\end{figure}

We demonstrate the performance of the Factorized Posterior (FP) approach with the EPTA 10-year dataset. 
We begin with the single pulsar analysis using the burst with displacement memory model to infer the intrinsic per-pulsar memory parameters. The resulting posteriors are then approximated using either KDEs or normalizing flows. 
The intrinsic memory parameters, such as per-pulsar strain amplitude and sign, are subsequently mapped to the global parameters defined by the memory burst’s strain amplitude, sky location, and polarization angle by equation~\eqref{eq:FP}. 
The factorized posterior is then constructed by evaluating the approximated per-pulsar densities (via KDE or flows) at these projected values and multiplying them across all pulsars, yielding an efficient approximation to the full-PTA likelihood in the global parameter space. 
Figure~\ref{fig:dr2new_fp_comparison} compares posteriors for the memory-burst parameters obtained with four methods on the EPTA 10-year data: full-PTA likelihood (blue), FP with lookup tables (LUT; green) \citep{SunBaker2023}, FP with kernel density estimation (KDE; red), and FP with normalizing flows (black). 

To summarize consistency between the FP approximations and the full-PTA posterior, we report two posterior agreement metrics.
First, the \emph{probability of superiority} (or common-language effect size) \(P(A>B)\) for each parameter, \textit{i.e.}, the probability that a random draw from the FP posterior exceeds one from the full-PTA posterior~\citep{mcgraw1992,hollander2013nonparametric}.
Values of \(P(A>B)\) near \(0.5\) indicate no systematic shift between the FP posterior and the full-PTA posterior.
Values about 0 and 1 point to a decrease or an increase of the average FP posterior sample value, respectively, compared to the full-PTA posterior.
Second, we introduce the \emph{fractional overlap of credible intervals} (CIO) at $1\sigma$ credibility as a simple heuristic for interval agreement~\citep{Inman01011989}. 
Larger CIO indicates stronger interval concordance.

We find the following probability of superiority and the CIO for our three FP posteriors against the full-PTA posterior.
For FP with KDE, we obtain the mean probability of superiority,
\(\overline{P(A>B)}=0.498\) (range \(0.418\)–\(0.529\) across parameters). 
For FP with normalizing flows, we obtain
\(\overline{P(A>B)}=0.498\) (range \(0.404\)–\(0.6\) across parameters). 
For FP with LUT, we obtain
\(\overline{P(A>B)}=0.499\) (range \(0.442\)–\(0.595\) across parameters). 
As for CIO, we find values in the range $(0.95$-$0.99)$ for all FP approaches and all parameters, except $\log_{10}A_\text{B}$, indicating an excellent agreement. 
For $\log_{10}A_\text{B}$, we find CIO in range ($0.3$-$0.4$), suggesting a significant difference in $1\sigma$ credible levels obtained with full-PTA analysis and the FP approach. 
We attribute this to difficulties in the approximation of the tail of the marginal posterior for $\log_{10}A_\text{B} < -13.5$. 
However, the peak of the marginalized full-PTA posterior matches well with those obtained with FP approaches. 
When calculating CIO for $3\sigma$ credible level for $\log_{10}A_\text{B}$, we find the values of $(0.92$-$0.95)$.
These numbers indicate close agreement of posterior approximations with the full-PTA result. 

While the full-PTA likelihood provides the most accurate inference, it comes with a high computational cost. The average time per likelihood evaluation is \(2.1984 \pm 0.0646\)~s for the full PTA likelihood, \(1.40 \pm 0.076\)~s for FP+KDE (\(\sim1.6\times\) faster), \(0.323 \pm 0.007\)~s for FP+flows (\(\sim6.8\times\) faster), and \(0.046 \pm 0.005\)~s for FP+LUT. Such improvements are crucial for enabling scalable Bayesian inference in high-dimensional PTA analyses. To reach this accuracy, LUTs require substantially more single-pulsar samples (i.e., a denser grid) to avoid interpolation artifacts and boundary bias. In contrast, flows and KDE yield continuous density approximations that remain accurate with fewer samples. LUTs offer extremely fast and deterministic evaluations but are discrete and memory-intensive. Flows/KDE provide smooth, differentiable densities that better capture curved posteriors and are easier to compose.

\label{sec:results:linear}

\section{Discussion}
\label{sec:discussion}

\paragraph{Sensitivity to Merging SMBHBs}

The analysis using SMBHB merger model builds on the framework established in the methods paper \citep{TomsonGoncharov2025}, which demonstrates that memory-inclusive SMBHB merger waveforms enable PTAs to probe a previously inaccessible region of the binary parameter sace - beyond the reach of existing continuous and memory burst models.

\paragraph{False positives.}
During the factorized posterior (FP) analyses, we have realized the importance of modeling the common-spectrum time-correlated stochastic process. 
Inclusion of this term is not straightforward for single-pulsar analyses from which the FP posterior is derived because two power law terms are degenerate in single-pulsar data. 
The workaround is to enforce a fixed spectral index in the common process term, as often done in searches for the gravitational wave background. 
Otherwise, we find false positives in our results. 
The same approach is used in~\citet{NG_15_MEM}. 
They have further adopted the model of pulsar-intrinsic time-correlated noise, where, instead of a power-law noise spectrum, noise power at every frequency is a free parameter. 
However, we find that it does not rule out the false positives. 

\paragraph{Posterior approximation.}
We note two general forms of mismatch that can arise in approximate posteriors: underconstraining, which typically leads to conservative but acceptable inference, and overconstraining, which can bias results and introduce false positives. In our analysis, the KDE approximation showed mild signs of overconstraining in some parameters, whereas the flow-based approximation remained robust. Overall, we find that both FP-based approximations reproduce the full-PTA likelihood posteriors reasonably well, with the flow model providing the closest match. Nevertheless, we recommend verifying the accuracy of the approximation in future analyses. Additionally, for the EPTA 10-year dataset using the SMBHB merger memory model, modelling HD correlations between pulsars is required. Factorized posterior approaches, which neglect these correlations, are therefore not suitable for searches that include correlations between the pulsars. Variational inference with normalizing flows, in contrast, is capable of capturing the full correlated posterior structure~\citep{VallisneriCrisostomi2025}.

KDE-based approximation required substantial empirical tuning to achieve comparable results. 
Due to the structured and sometimes multimodal nature of the posterior in memory burst searches, naive KDE settings led to either over-smoothed distributions or spurious modes.
Through iterative refinement, we found that a relatively small kernel bandwidth of $0.05$ provided the best compromise, accurately capturing sharp features while minimizing artifacts. This highlights a limitation of KDE in approximating complex posteriors.


\section{Conclusion}
\label{sec:conclusion}

We performed a search for a number of gravitational wave displacement memory signals. 
First, we developed a search for SMBHB mergers, including the signal model with null memory.
We performed the search on PPTA DR3, 10-year and 25-year EPTA DR2 datasets. 
Second, we searched for generic bursts of displacement memory in all three datasets.
We placed upper limits on the strain amplitude of memory bursts, ruling out the observation of these signals with strain amplitude $h_0 > 10^{-13}$ over the period of $22$ years at $95\%$ credibility. 
We rule out the existence of generic memory bursts with amplitudes $h_0 > 10^{-14}$ in brief periods of the observation time. 
With the full NR model for SMBHB mergers, we rule out the existence of null memory bursts with $h_0 > 10^{-14}$ over a longer period of about $7$ years.
We have also placed lower limits on the luminosity distance $D_\text{L}$ for SMBHBs with $\mathcal{M}$ between $10^8$ and $10^{10}~M_\odot$ at $95\%$ credibility. 
With a 25-year EPTA DR2, we rule out the mergers of SMBHBs with $\mathcal{M}=10^{10}~M_\odot$ up to $280$~Mpc.
With PPTA DR3 spanning $18$ years, we rule out the mergers of SMBHBs with $\mathcal{M}=10^{10}~M_\odot$ up to $700$~Mpc.

\section{Acknowledgements}
We are grateful to Abhimanyu Sushobhanan and Wang Wei Yu for their insightful scientific discussions and support with data analysis throughout the course of this project. We also thank Bruce Allen for helpful comments. This work was supported by the Max Planck Gesellschaft (MPG) and the ATLAS cluster computing team at AEI Hannover.

We acknowledge the use of data from the Parkes Pulsar Timing Array (PPTA), and support from the Australian Research Council Centre of Excellence for Gravitational-Wave Discovery (OzGrav; Grants CE170100004, CE170100007, CE23010016). 
We also extend our gratitude to the European Pulsar Timing Array (EPTA) and the facilities that provide its data: the 100-m radio telescope in Effelsberg operated by the Max-Planck-Institut für Radioastronomie (MPIfR), the Westerbork Synthesis Radio Telescope (operated by ASTRON, with support from the Netherlands Foundation for Scientific Research, NWO), the Nançay Radio Observatory (operated by Paris Observatory under CNRS and Université d’Orléans, supported through PNCG and PNHE), and the Sardinia Radio Telescope (funded by MIUR, ASI, and RAS; operated by INAF). We acknowledge support from GRAvity from Astrophysical to Microscopic Scales (GRAMS, ERC Consolidator Grant No. 815673), the ERC Synergy Grant “GWSky” (Grant No. 101167314), the PRIN 2022 project GUVIRP, and the Marie Sklodowska-Curie Fellowship No. 101007855.

\software{
\textsc{ptmcmcsampler}~\citep{EllisvanHaasteren2019}, 
\textsc{enterprise}~\citep{EllisVallisneri2020},  
\textsc{floz}~\citep{SrinivasanCrisostomi2024} at \href{https://github.com/Rahul-Srinivasan/floZ}{github.com/Rahul-Srinivasan/floZ}, 
\textsc{gwsurrogate}~\citep{FieldVarma2025}, 
\textsc{enterprise\_warp} at \href{https://github.com/bvgoncharov/enterprise_warp}{github.com/bvgoncharov/enterprise\_warp}. 
}

\bibliography{mybib,collab}

\appendix

\section{The results of parameter estimation}
\label{sec:appendix:pe}

\begin{figure}[htbp]
  \centering
  \includegraphics[width=0.6\textwidth]{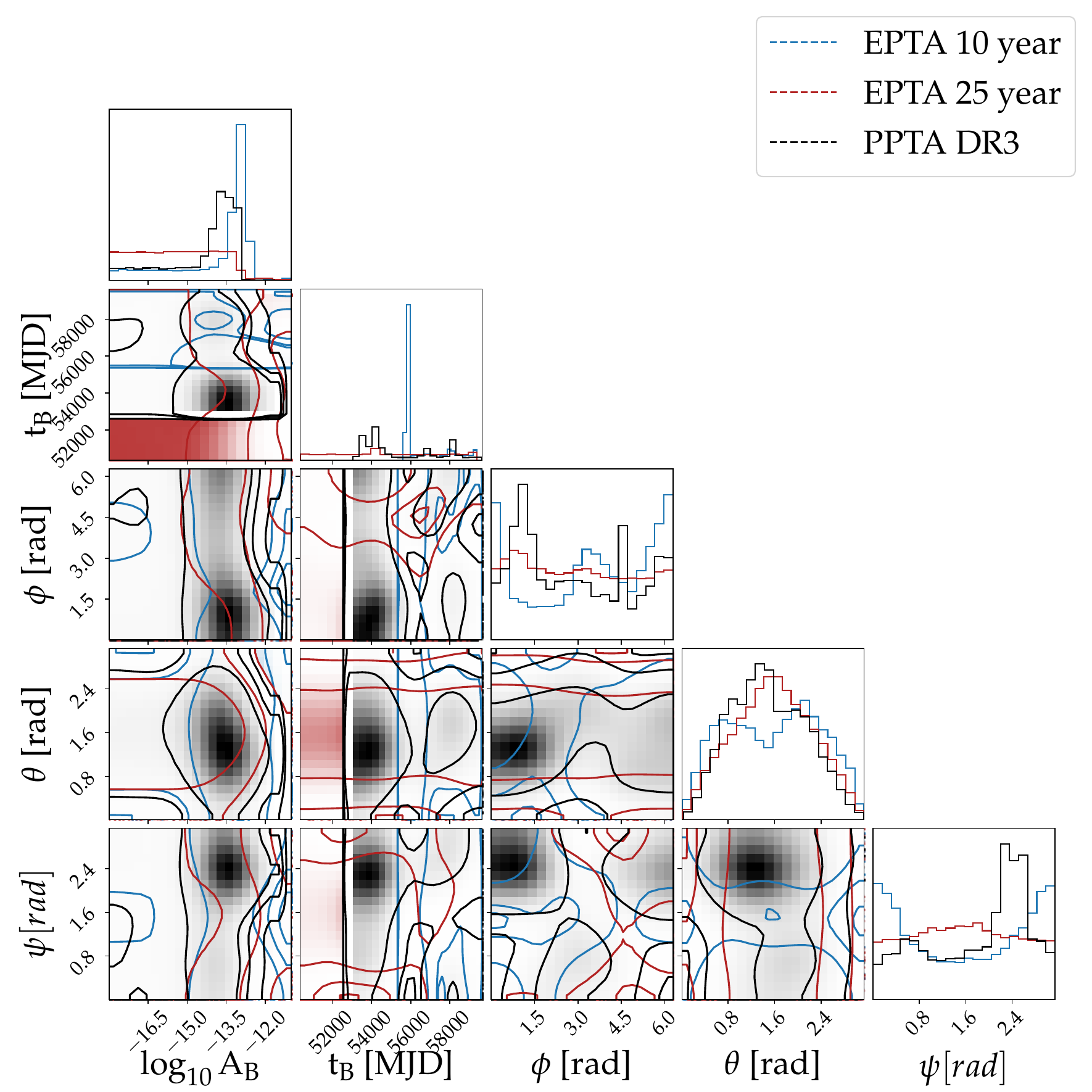}  
  \caption{
  Posterior distributions of gravitational wave memory signal parameters from generic memory burst search on 10 year EPTA (blue), 25 year EPTA (red) and PPTA DR3 (black) datasets. The prior ranges on the burst epoch \( t_0 \)[MJD] for each dataset are: \( \mathcal{U}(55611,\,59385) \) for 10-year EPTA, \( \mathcal{U}(50360,\,59385) \) for 25-year EPTA, and \( \mathcal{U}(50340,\,59640) \) for PPTA DR3.
  }
  \label{fig:burst_post}
\end{figure}
\begin{figure*}[htbp]
  \centering
  \gridline{
    \fig{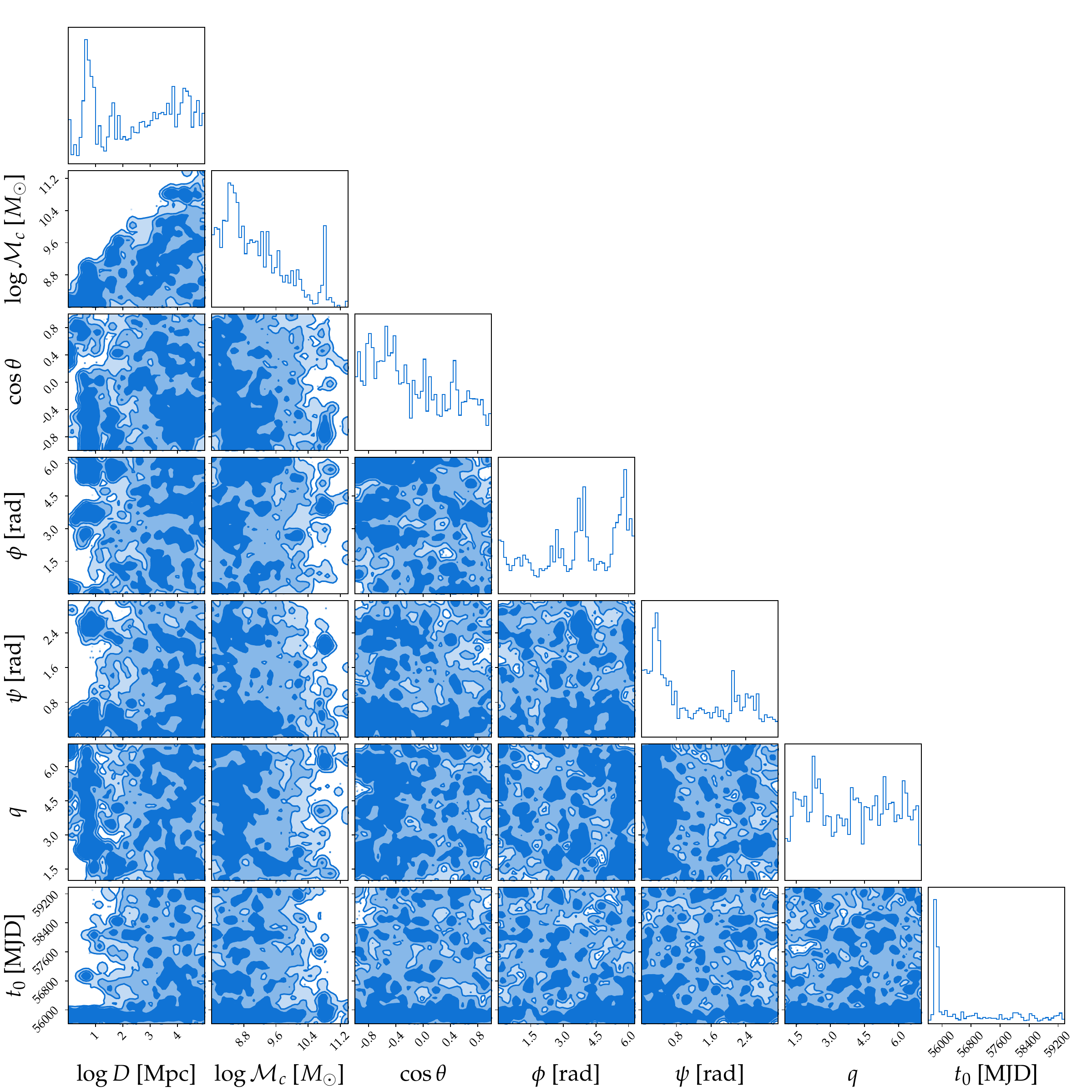}{0.49\textwidth}
        {(a) 10-year EPTA DR2\label{fig:mem_post:epta10}}
    \fig{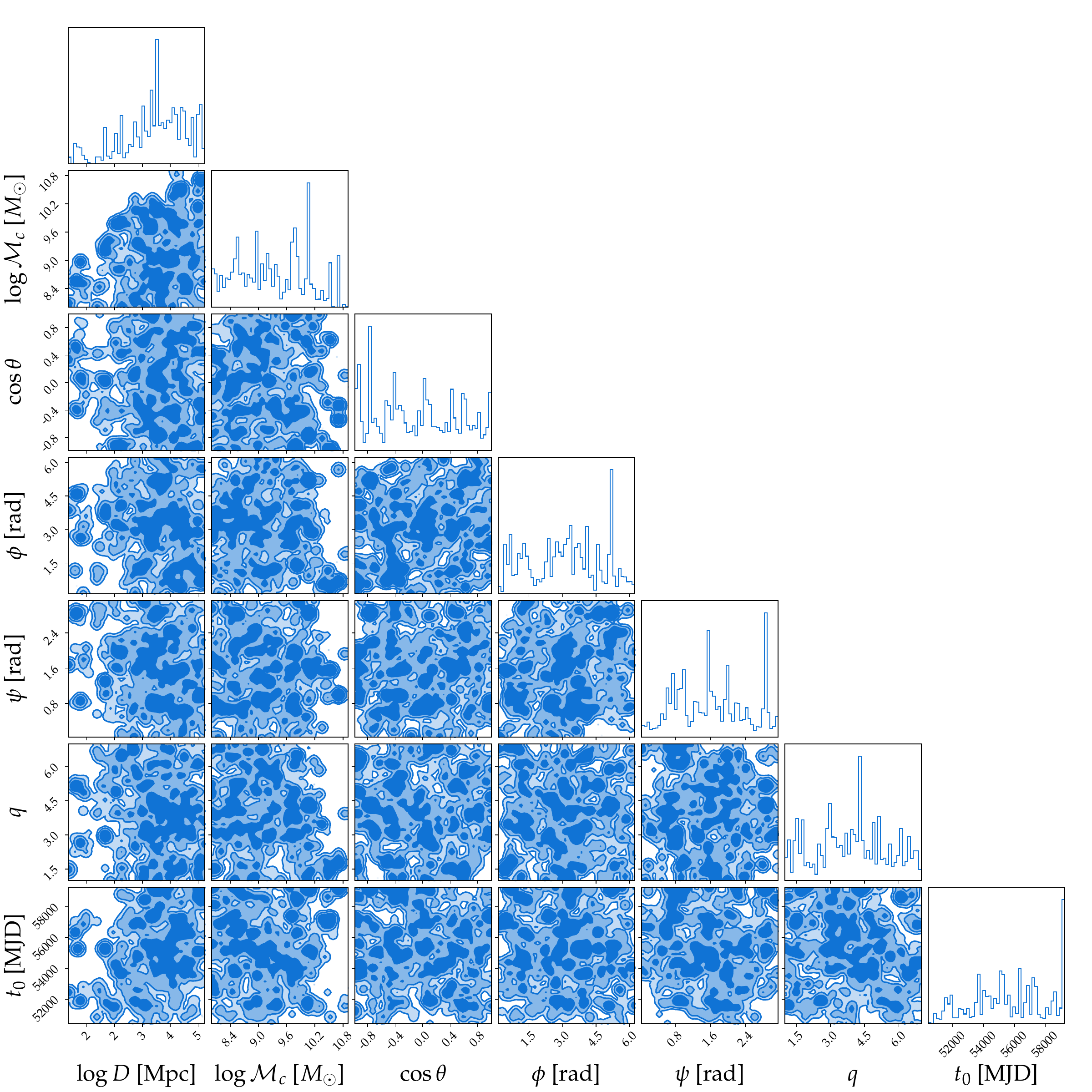}{0.49\textwidth}
        {(b) 25-year EPTA DR2\label{fig:mem_post:epta25}}
  }

      \caption{Posterior distributions for search for SMBHB merger with null memory.  
  Parameters are chirp mass $\mathcal{M}$, mass ratio $q$, luminosity distance $D_\text{L}$, burst epoch $t_\text{B}$, right ascension, declination, and polarization angle $\psi$ of the expected SMBHB merger.  
  Panel (a) shows results for the 10-year EPTA DR2 data set; panel (b)  25-year EPTA DR2 data}
  \label{fig:mem_post}
\end{figure*}
\begin{figure}
    \centering
    \includegraphics[width=0.5\linewidth]{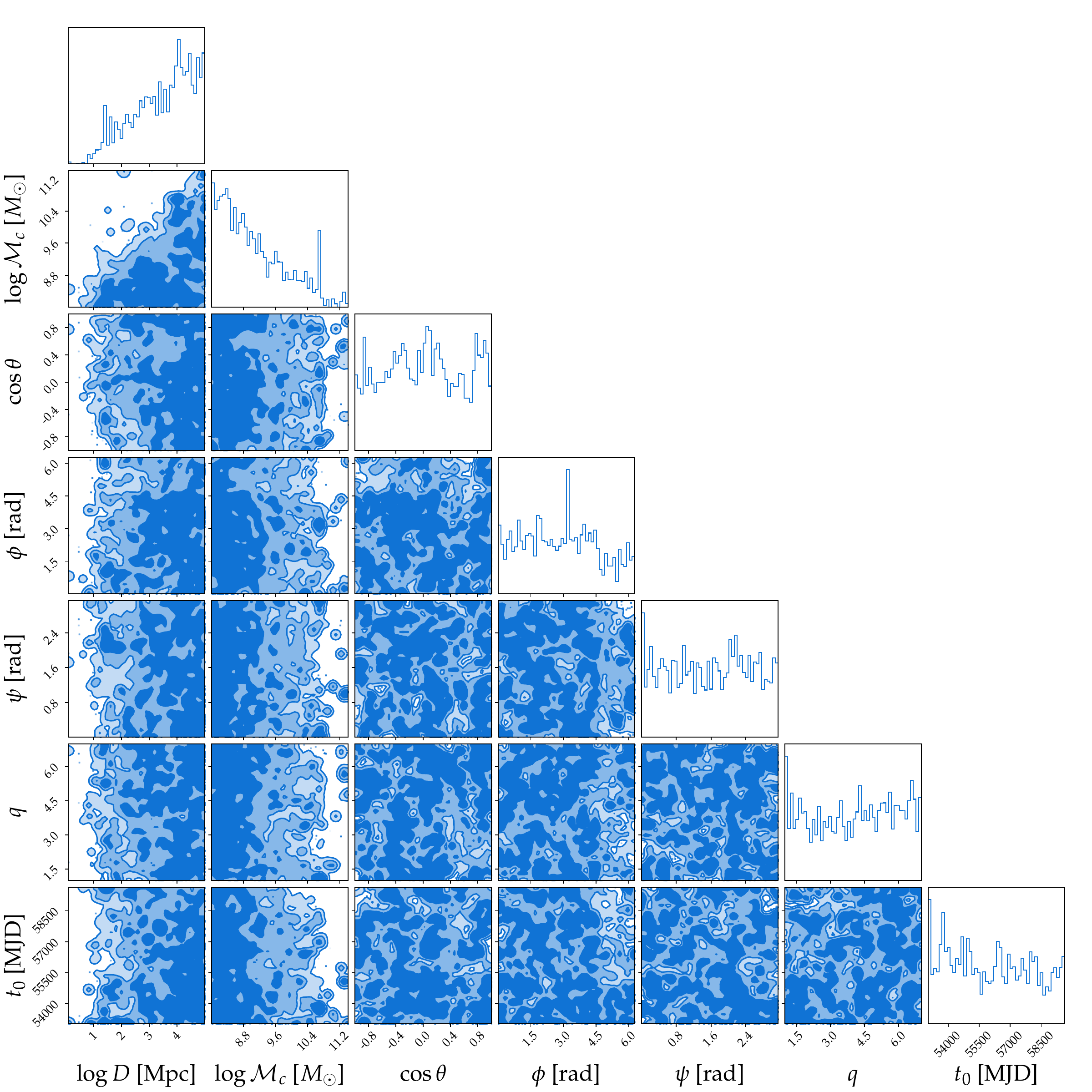}
    \caption{Posterior distributions for search for SMBHB merger with null memory on PPTA DR3 dataset}
    \label{fig:mem_post_ppta}
\end{figure}

In this Section, we show the results of the estimation of all parameters for the models and the data we discussed in this work. Figure~\ref{fig:burst_post} depicts the posteriors of the generic memory burst search on EPTA 10 year, 25 year datasets and the PPTA DR3 datasets. In Figure~\ref{fig:mem_post} and~\ref{fig:mem_post_ppta}, we show the posterior for all parameters of the NR waveform model in a search for SMBHB mergers with null memory. Our priors are shown in Table~\ref{tab:prior_burst}.

\section{Comparison of Posteriors across Methods}
\label{sec:appendix:fp}

\begin{figure}[htbp]
  \centering
  \includegraphics[width=0.5\textwidth]{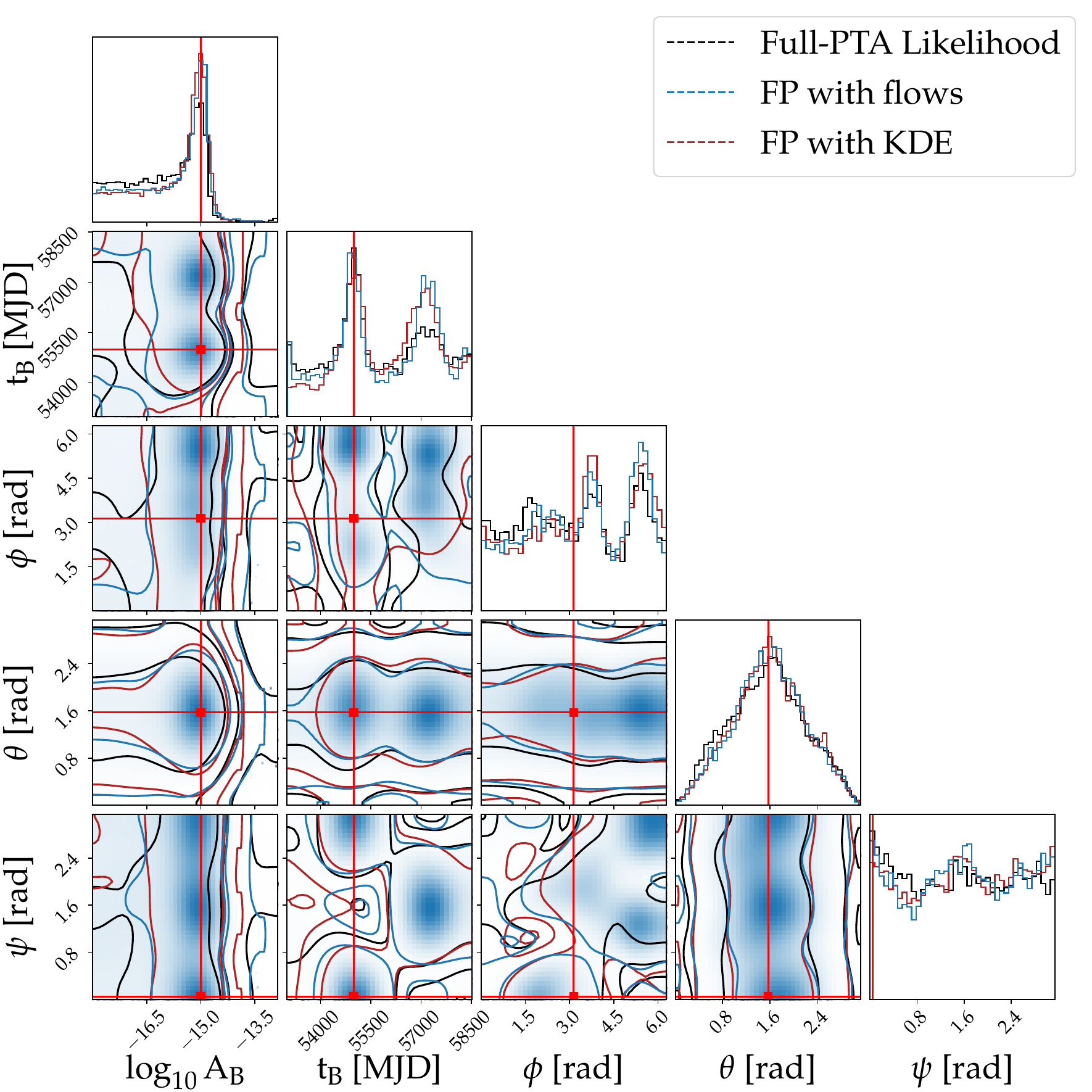}  
  \caption{
  Comparison of posterior distributions for a simulated memory burst signal obtained using three inference approaches: (i) the full-PTA likelihood computed with the standard enterprise framework (black), (ii) the Factorized Posterior (FP) approach with kernel density estimation (KDE, red), and (iii) the FP approach using normalizing flows (blue).
  Vertical and horizontal red lines show simulated parameter values. 
  }
  \label{fig:Full_array_vs_KDE_vs_nflows}
\end{figure}

In this Section, we compare the posterior distributions obtained using three different approaches : Full-PTA Likelihood, Factorized Posterior (FP) Approach using Kernel Density Estimates (KDE), and FP using normalizing flows. Figure~\ref{fig:Full_array_vs_KDE_vs_nflows} depicts the comparison of the distribution for a simulated dataset consisting of 10 pulsars with an simulated generic burst memory. The simulated (true) burst parameters are depicted as vertical and horizontal lines in the figure. As shown in Figure~\ref{fig:Full_array_vs_KDE_vs_nflows}, the results from the flow-based and KDE approximations are generally consistent with the full PTA result, capturing the overall shape and covariance of the distribution. 

The choice of bandwidth in KDE plays a crucial role in balancing bias and variance—an overly large bandwidth can over smooth features, while a small one may lead to noise amplification. Similarly, for flow-based models, the number of flow layers, hidden units, and coupling block structure significantly affect the expressiveness and convergence of the model. Careful selection and tuning of these parameters is necessary to ensure accurate density approximation, particularly in multimodal distributions. Flow-based models are particularly well-suited for capturing multimodal distributions due to their ability to learn complex, nonlinear transformations of the posterior.

\section{Single Pulsar Limits}
\label{sec:appendix:singlepsr}

\begin{figure*}[htbp]
  \centering
  \gridline{
    \fig{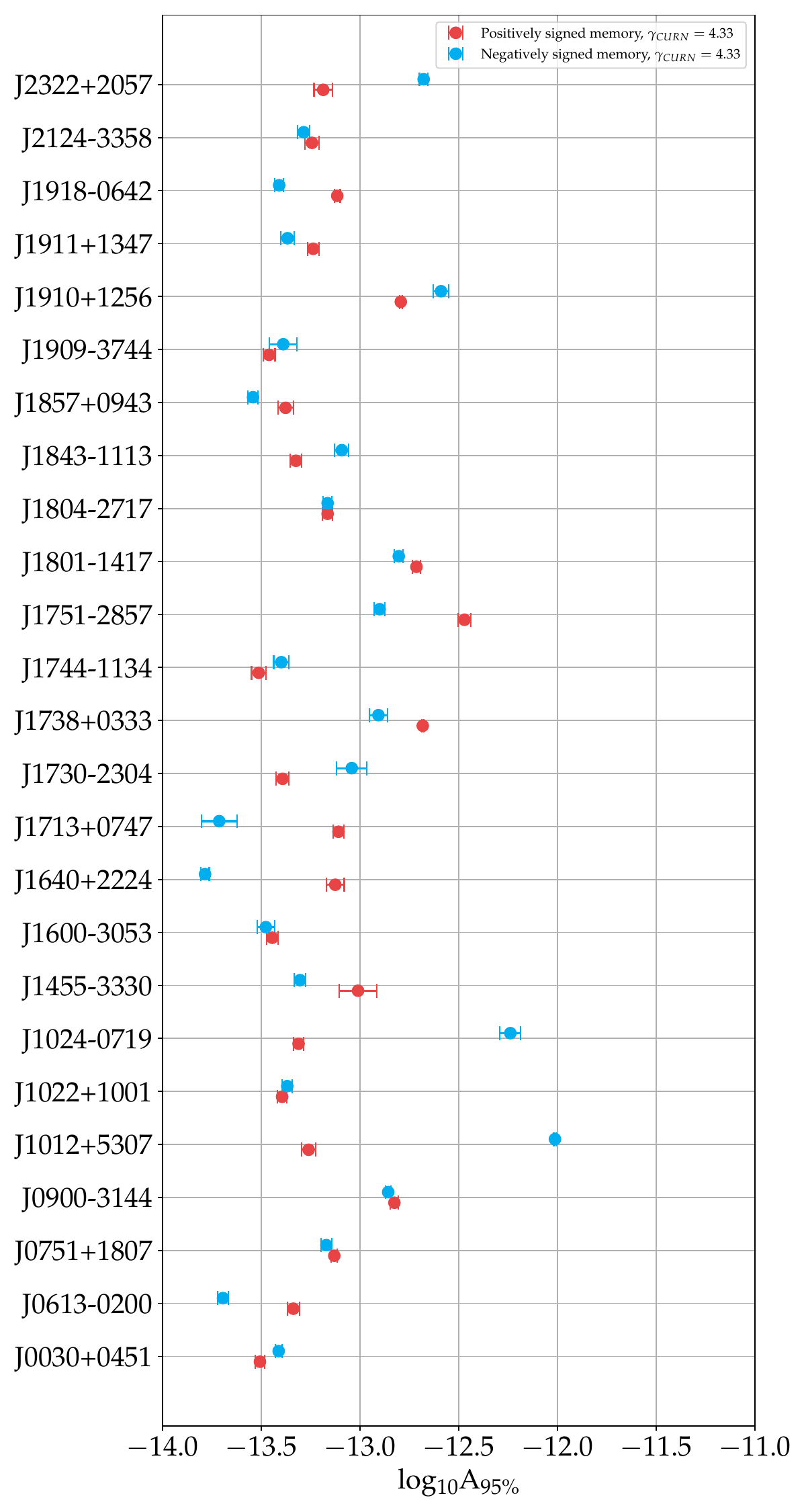}{0.32\textwidth}
        {(a)\label{fig:pul_ul:epta10}}
    \fig{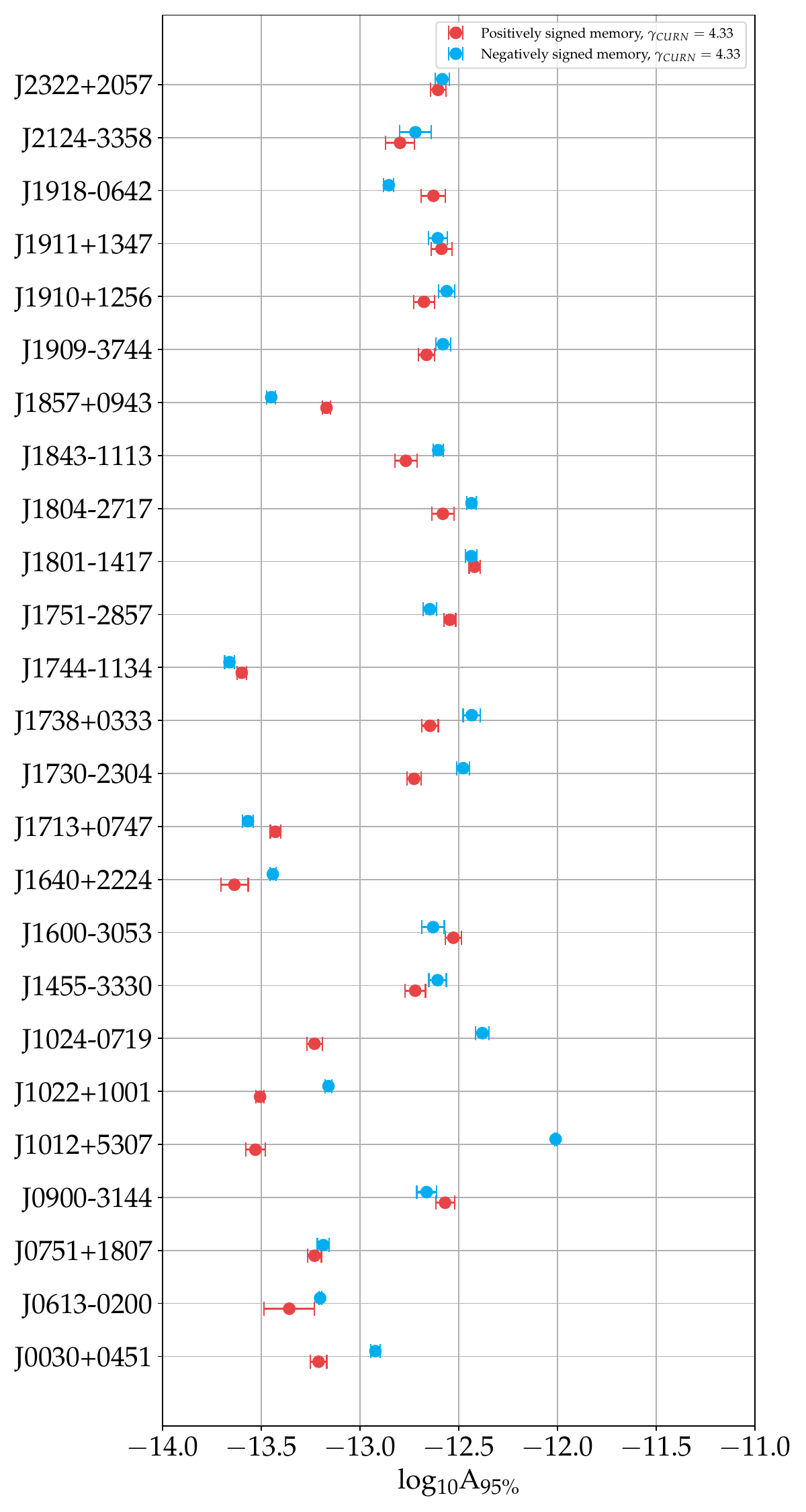}{0.32\textwidth}
        {(b)\label{fig:pul_ul:epta15}}
    \fig{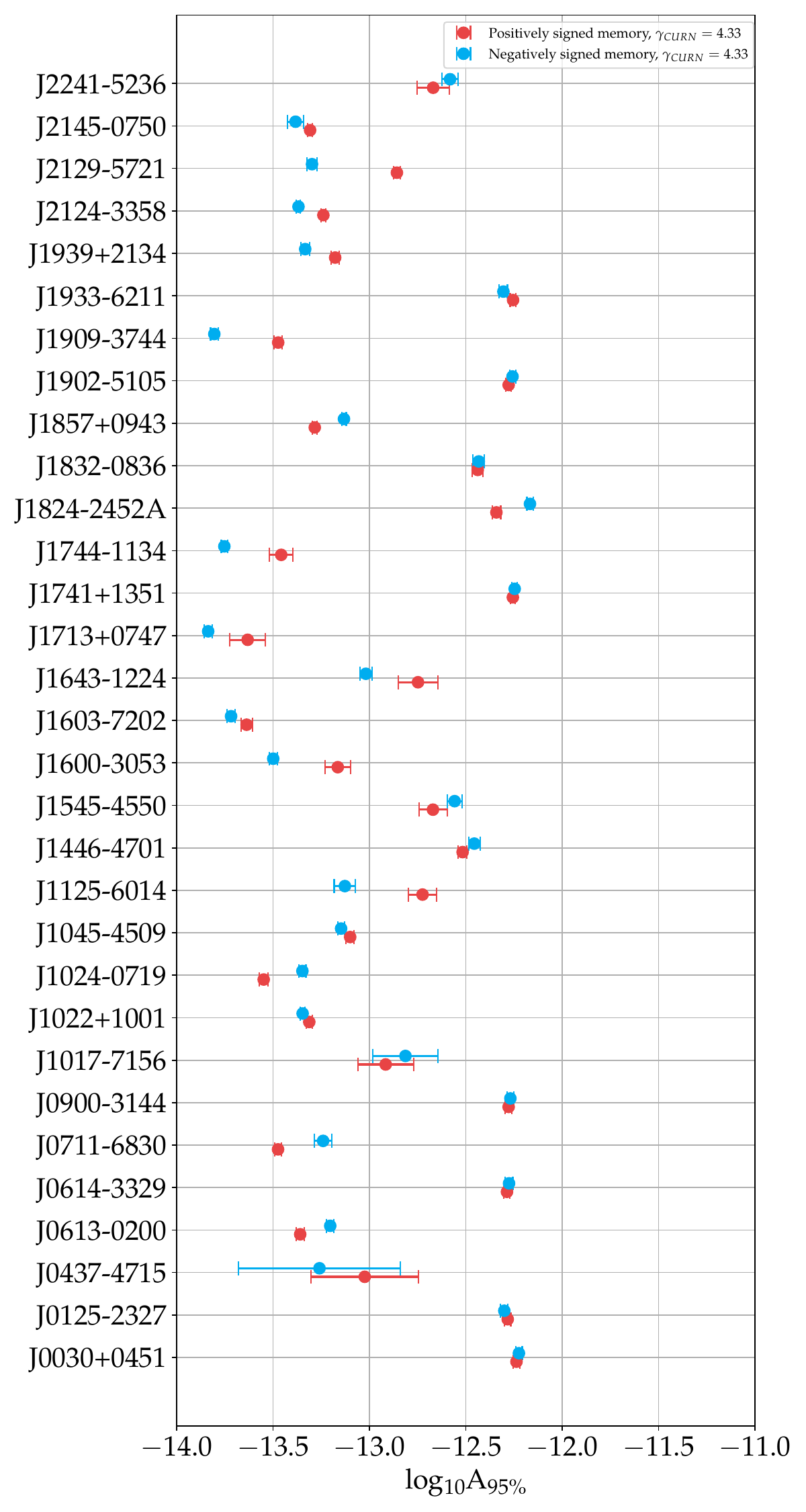}{0.32\textwidth}
        {(c)\label{fig:pul_ul:ppta}}
  }

  \caption{Pulsar-term upper limits on the strain amplitude $h_{0}$, marginalising over burst epoch as well as each pulsar’s red‐ and white-noise parameters.  
  Panels (a)–(c) show results for the EPTA 10-year, EPTA 15-year, and PPTA DR3 data sets, respectively (left → right).}
  \label{fig:pulsar_term_UL}
\end{figure*}

We show single-pulsar limits on the memory burst amplitude for 10-year EPTA DR2, 25-year EPTA DR2, and PPTA DR3 in Figure~\ref{fig:pulsar_term_UL}.

\begin{table}[ht]
  \centering
  \setlength{\tabcolsep}{6pt}   
  \renewcommand{\arraystretch}{1.3}   
  \begin{tabular}{|l|c|c|}
    \hline
    \textbf{Parameter} & \textbf{Prior} & \textbf{Description} \\ \hline\hline
    \multicolumn{3}{|c|}{Memory burst model}  \\ \hline
    $\log_{10}h_{0}$   & $\mathcal{U}(-17,\,-10)$ & Log–10 amplitude  \\ \hline
    \multicolumn{3}{|c|}{NR waveform model (SMBHB mergers)}  \\ \hline

    $ \log_{10}\mathcal{M}$~[$M_\odot$]   & $\mathcal{U}(8,\,12)$ & Log-10 Chirp Mass  \\ \hline
    $\log_{10}D_{L}$~[Mpc]   & $\mathcal{U}(0,\,5)$ & Log-10 Luminosity Distance  \\ \hline
    $q$   & $\mathcal{U}(1,\,7)$ & Mass ratio  \\ \hline
    \multicolumn{3}{|c|}{All models}  \\ \hline
    $\psi$             & $\mathcal{U}(0,\,\pi)$   & Polarization      \\
    $\cos\theta$           & $\mathcal{U}(-1,\,1)$   & Cosine of Polar angle       \\
    $\phi$             & $\mathcal{U}(0,\,2\pi)$  & Azimuthal angle   \\
    $t_{0}$~[MJD]            & $\mathcal{U}(55611,\,59385)$  & 10-year EPTA DR2 \\
    ~                  & $\mathcal{U}(50360,\,59385)$ & 25-year EPTA DR2 \\
    ~                  & $\mathcal{U}(50340,\,59640)$ & PPTA DR3         \\ \hline
  \end{tabular}
  \caption{Priors used in our analysis.}
  \label{tab:prior_burst}
\end{table}

\end{document}
